   \documentclass{aa} 

\usepackage{graphicx}
\usepackage{txfonts}

\graphicspath{{Figures/}}

 \newcommand{\s}{\nobreak\hspace{.11em}\nobreak}

  \newcommand{\be}{\begin{equation}}
 \newcommand{\ee}{\end{equation}}
 \newcommand{\ba}{\begin{eqnarray}}
 \newcommand{\ea}{\end{eqnarray}}
 \newcommand{\bs}{\begin{subequations}}
 \newcommand{\es}{\end{subequations}}

\begin{document}

   \title{Pathways of survival for exomoons and inner exoplanets}


   \author{Valeri V. Makarov
          \inst{1}
          \and
          Michael Efroimsky\inst{2}
          }

   \institute{US Naval Observatory, 3450 Massachusetts Avenue NW, Washington DC 20392 USA\\
              \email{valeri.makarov@gmail.com}
         \and
             US Naval Observatory, 3450 Massachusetts Avenue NW, Washington DC 20392 USA\\
             \email{michael.efroimsky@gmail.com}
             }

   \date{Received November 22, 2022; accepted February 9, 2023}

 \authorrunning{Valeri V.Makarov and Michael Efroimsky}
 \titlerunning{Pathways of survival for exomoons and inner exoplanets}

  \abstract
   {It is conceivable that a few thousand confirmed exoplanets initially harboured satellites similar to the moons of the Solar System or larger. We ask the question of whether some of them have survived over the {\ae}ons of dynamical evolution to the present day. The dynamical conditions are harsh for exomoons in such systems because of the greater influence of the host star and of the tidal torque it exerts on the planet.}
   {We investigate the stability niches of exomoons around
   { hundreds of innermost exoplanets for which the needed parameters are known today,} and we determine the conditions of these moons' long-term survival.
  {General lower and upper bounds on the exomoon survival niches are derived for orbital separations, periods, and masses.}}
   {The fate of an exomoon residing in a
stability niche depends on the initial relative rate of the planet’s rotation and
on the ability of the moon to synchronise the planet by overpowering the
tidal action from the star. State-of-the-art models of tidal dissipation and secular orbital evolution are applied to a large sample of known exoplanet systems, which have the required estimated physical parameters.}
   {We show that in some plausible scenarios, exomoons can prevent close exoplanets from spiralling into their host stars, thus extending these planets’ lifetimes. This is achieved when exomoons synchronise the rotation of their parent planets, overpowering the tidal action from the stars.}
   {Massive moons are more likely to survive and help their host planets  maintain a high rotation rate  (higher than these planets’ mean motion).}


   \maketitle
%

 \section{Introduction\label{Section 1}}

 A common convention regarding the rotation of close-in exoplanets is that their tidal de-spinning is fully defined by the tides generated on these planets by their host stars.
 It is for this reason that close-in gas giants are assumed to be synchronised or pseudo-synchronised. We pose the question of whether a moon orbiting a planet can change the outcome of the planet's tidal de-spinning and make the planet end up in a non-synchronous (perhaps, even retrograde) spin state. Specifically, we explore the possibility of a planet being synchronised not by the star but by a moon. If this scenario is plausible for some inner planets, these planets' rotation will be much faster than the speed at which they orbit the star. These planets then avoid spiraling into their host stars, and maintain finite orbital eccentricities.\,\footnote{~As follows from the first line of Eq. (143) in \citet{BoueEfroimsky},
 a rotation rate faster than the mean motion works to tidally increase the semi-major axis in the two-body problem. Also, according to their equations (156 -157),
 fast rotators tend to boost the eccentricity.}  An internally synchronised planet--moon pair continues to shrink because of the tides caused by the star, but the rate of this process is orders of magnitude slower than the orbital decay of a moon-less planet. Indeed, in an
 {internally} synchronised or pseudo-synchronised planet-moon system, the tidal torque acting on the planet from the star is counterbalanced by the tidal torque acting on the planet from the moon. The time-averaged total torque on the planet in a stable equilibrium is nil, but both these components continue to dissipate energy. This energy is taken from the orbital kinetic energy of the planet and the moon, respectively. The supply of orbital energy is orders of magnitude greater than the rotational energy of the planet, which makes the decay process very slow. A decisive test of these models will be possible once more accurate photometric measurements of the secondary eclipses of hot Jupiters become available (i.e. the transits of the planets behind the disks of the stars in the upper conjunction), where a non-synchronous rotation is expected to produce a measurable phase shift with the peak brightness of the out-of-eclipse light curve. A statistically significant shift towards the evening terminator was detected for the WASP-12b planet by \citet{2021MNRAS.503L..38O}. On a fast-rotating planet, the hot spot generated by the stellar irradiation is not stationary but rather leads the substellar meridian.

 The dynamical evolution of exomoons in star-planet systems has been investigated in a number of publications \citep[e.g.,][]{Sasaki,2002ApJ...575.1087B, 2018AJ....156...50G}. The commonly shared conclusion is that the chances of exomoons surviving for an extended duration of time in known planetary systems are low. Considering three basic scenarios for a moon in such systems, with only one of these scenarios allowing the moon to survive for at least 1 Gyr, \citet{2021PASP..133i4401D} conclude that the survival rate is close to nil for planets with orbital periods of 10 d or less, and that it gradually increases to 70\% for periods of 300 d. The median orbital period of all detected and
 \begin{table*}[h]
 \begin{minipage}{190mm}
  \caption{Symbol key}\vspace{6mm}
  \label{description}
  \begin{tabular}{@{}lll@{}}
  \hline
   Notation & Explanation & Reference \\
 \hline

$M_*$ & mass of the star &   \\[2pt]
 $M_p$ & mass of the planet &   \\[2pt]
 $M_m$ & mass of the moon &   \\[2pt]
 $\rho_m$ & density of the moon &   \\[2pt]
 $R_p$ & radius of the planet &   \\[2pt]
 $R_m$ & radius of the moon &   \\[2pt]

   $C_p$ & maximal moment of inertia of the planet &  \\[2pt]

   $C_m$ & maximal moment of inertia of the moon &  \\[2pt]

 $\theta_p$          &  rotation angle of the planet    &             \\[2pt]


 $a_p$ & semi-major axis of the planet &    \\[2pt]

 $a_m$ & semi-major axis of the moon &    \\[2pt]

 $e_p$ & eccentricity of the planet &     \\[2pt]

 $e_m$ & eccentricity of the moon &     \\[2pt]



 $n_p$ & mean motion of the planet&   \\[2pt]

 $n_m$ & mean motion of the moon &   \\[2pt]

 $lmpq$ & integers used to number the tidal Fourier modes &  eqn (\ref{gedo}) \\[2pt]

 $F_{lmp}(i)$ & inclination functions & eqn (\ref{gedo})   \\[2pt]
 $G_{lpq}(e)$ & eccentricity function &  eqn (\ref{gedo})  \\[2pt]

 $\omega_{lmpq}$ & Fourier modes of the tides in the planet & eqn (\ref{omega}) \\[2pt]

 $\chi_{lmpq}\,\equiv\,\mid\omega_{lmpq}\mid\,$& forcing frequencies excited in the planet & eqn (\ref{chi})    \\[2pt]

 $\epsilon_l\,=\,\epsilon_l(\omega_{lmpq})$ & tidal phase lags in the planet &  \\[2pt]

 $k_l\,=\,k_l(\omega_{lmpq})$ & dynamical Love numbers of the planet &   \\[2pt]

 $K_l(\omega_{lmpq})= k_l(\omega_{lmpq})\sin\epsilon_l(\omega_{lmpq})$ & quality functions of the planet &  eqn (\ref{notation}) \\[2pt]

 $ Q_l( \omega_{lmpq} )\,\equiv\,{1}\s/\s{\textstyle{\mid\sin\epsilon_l(\omega_{lmpq})\mid}}$ &  tidal quality factors   & eqn (\ref{q})\\[2pt]

$r_H$ & Hill radius of the planet & eqn (\ref{rH.eq})\\[2pt]

$r^{\,\prime}_H$ & reduced Hill radius of the planet & eqn (\ref{8})\\[2pt]

 $r_R$ & Roche radius of the planet & eqn (\ref{1})\\[2pt]

 $a_m^{(s)}$ & synchronous value of the lunar semi-major axis  & eqn (\ref{2}) \\[2pt]

${\cal T}^{(\ast)}$ & solar polar tidal torque acting on the planet & \\[2pt]

${\cal T}^{(m)}$ & lunar polar tidal torque acting on the planet & \\[2pt]

 $G$ & Newton's gravitational constant & \\[2pt]
\hline
\end{tabular}
\end{minipage}
\end{table*}

\noindent
tentative exoplanets is about 11.8 d, and only 10\% of them have periods above 442 d,
so we should expect to find
very few exoplanets with moons.

 In the current work, our goal is to demonstrate that there are additional pathways for exomoons to survive, even around fairly close-in planets. Furthermore, exomoons can help their parent planets survive for longer times in the close vicinity of stars. As an example of such previously ignored possibilities, moons that are initially on retrograde orbits with respect to the spin of a slowly rotating planet will generate a tidal torque on the planet that will be opposed to the torque caused by the star.
  {Possible in principle,} this particular scenario is unlikely to play a large role in real systems, because the capture of external moons onto retrograde orbits is relatively rare.\,\footnote{~Two capture mechanisms are known to permit the acquisition of not only prograde but also retrograde moons: a binary-planet gravitational encounter \citep{Agnor2006} and chaos-assisted capture \citep{AstakhovFarrelli2004,Astakhov}. The latter mechanism was studied for small irregular moons, though in principle it may facilitate a capture of a larger body.
 ~\\
 The possibility of moons' inclination flips under close encounters is still awaiting exploration. For comets and asteroids, such flips are definitely possible \citep{Valsecchi}.} Starting with solid principles of tidal dynamics, we aim to show the existence of mechanisms plausible for a long-term coexistence of moons with their planets (most of the considered planets being dangerously close to their stars). Along with Earths and super-Earths, our study addresses Jupiters. It is commonly assumed that close-in low-viscosity giant planets are synchronised by their host stars. Although the tidal quality of Jupiters is likely to be as high as that of our own Jupiter \citep{2022Univ....8..211E}, the dissipation of orbital energy inside a synchronised planet is almost nullified because of the circularisation process bringing down the planet's eccentricity. The fate of the planet is then defined by the rotation rate of the host star, which is often slower than the mean motion. Slowly but relentlessly, the tidal break will remove the orbital energy, and the planet will plunge into the star. We however show here that a sufficiently massive moon can arrest or even reverse this process for a significantly long time, until the moon's orbital energy is depleted.

  \section{The Hill sphere and the reduced Hill sphere\label{hill.sec}}
 ~\\
 The Hill sphere is a region where the motion of moons is defined predominantly by the gravity of the planet, and less by that of the star.
 Analysis by \citeauthor{Hamilton} (\citeyear{Hamilton})
set its radius at
 \ba
 r_H\,=\;a_p\;(1-e_p)\,\left(\frac{M_p}{3\,M_{\ast}} \right)^{1/3}\;\,,
 \label{rH.eq}
 \ea
 where $a_p$ and $e_p$ are the planet's semi-major axis and eccentricity, while $M_p$ and $M_{\ast}$ are the masses of the planet and the star, correspondingly.

   However, numerical analysis by \citeauthor{Astakhov} (\citeyear{Astakhov}) and \citeauthor{Domingos}
(\citeyear{Domingos})
indicated that the orbits too closely approaching this radius become unstable in the long term, while those confined to a smaller domain (which we term {{`the reduced Hill sphere'}}) remain stable. The radius of the reduced Hill sphere (`the reduced Hill radius') for a circular ($\s e_m=0\s$) orbit of the moon is
   \ba
   r^{\,'}_H\,=\,B\;a_p\;(1\,-\,b\;e_p)\,\left(\frac{M_p}{3\,M_{\ast}} \right)^{1/3}
   \label{8}
   \ea
   and depends on the direction of the orbit:
   \bs
   \ba
   B\;=\;0.49\;,\;\;b\,=1.03\;\,,\;\;\;\s\mbox{for prograde-orbiting moons}\s;\,
   \label{9a}
   \ea
   \ba
   B\;=\;0.93\;,\;\;b\,=1.08\;\,,\;\;\;\,\mbox{for retrograde-orbiting moons}\s.
   \label{9b}
   \ea
   \label{9}
   \es


   \begin{figure}
   \resizebox{\hsize}{!}
         {\includegraphics 
         {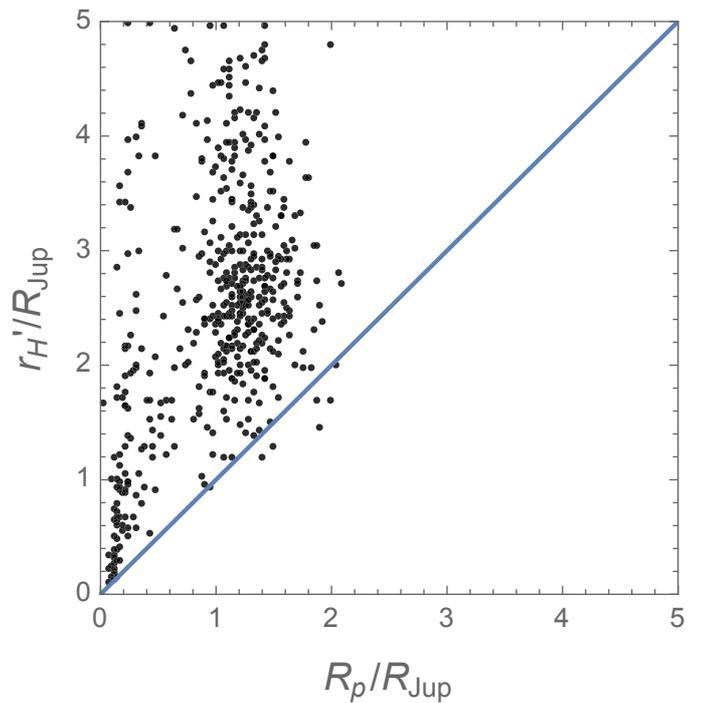}}
      \caption{Reduced Hill radii of 580 known inner exoplanets versus their radii in units of Jupiter's radius. The diagonal line indicates the equality of the two radii.}
         \label{rhill.fig}
   \end{figure}

 Figure  \ref{rhill.fig} shows the distribution of calculated reduced Hill radii for a sample of 580 inner exoplanets with the necessary estimated parameters entering Eq. (\ref{8}) provided in the
  {NASA Exoplanet Archive \footnote{\url{https://exoplanetarchive.ipac.caltech.edu/docs/data.html}
  ~~~~The version as of December 2022.}} (assuming prograde orbits). These values are plotted against the estimated radii of the planets themselves. One should expect the Hill spheres to be outside of the planets' radii, but this seems to not always be the case. The planets orbiting HAT-P-32, HAT-P-65, HATS-27, HATS-69, HIP 65 A, WASP-121, and WASP-19
   {appear to be} in violation of this condition. These are bloated Jupiters with orbital periods between 0.79 and 4.64 d orbiting solar-type stars.
   {However, caution should be exercised with the archival data for specific exoplanets, where generous upper limits are sometimes given for orbital eccentricity instead of true determinations. This concerns the planets HAT-P-65, HATS-27, and HATS-69. Still, the existing models suggest that these giant planets, being so close to the Hill radii, } should be shedding their outer layers, probably forming a gaseous disk around the stars. The origin of such critical proximity of inflated hot Jupiters to their host stars is not known.

 Equations (\ref{8} - \ref{9}) also yield
 an inequality to be utilised below:
 \ba
 a_m\s <\s r_H^{\,\prime}\s\approx\s B\s a_p\left( \frac{\textstyle M_p}{\textstyle 3\s M_*} \right)^{1/3}\;\,,
 %
 \label{inequality}
   \ea
  {$a_m$ and $a_p$ being the semi-major axes of the moon and the planet.}

 \section{The Roche radius \label{Section 2.1}}

 The Roche radius (or the Roche limit) is the closest distance at which a moon can get to its planet without being pulled apart by tides. This parameter sets the lower limit on the survival niche for exomoons.
 For a spherical host, the Roche radius is given by
 \ba
 r_R\,
  =\;A\,R_m\left(\frac{M_p}{M_m  }\right)^{1/3}\,\;,
 \label{}
 \ea
 where $M_m$ and $R_m$ are the mass and radius of the moon, while $M_p$ is the mass of the planet.

 For a strength-less moon, the dimensionless parameter would assume the value $A \approx 2.46$ borrowed from a calculation by \citet{Chandrasekhar} for an  incompressible fluid body. The shear strength of realistic rubble piles enables them, however, to survive at smaller radii, with $A\approx2.2\,$  (\citeauthor{Leinhardt} \citeyear{Leinhardt}).\footnote{~For porous aggregates stronger than rubble but weaker than solid, like Phobos, the presence of tensile strength renders the values of $A$ that may, probably, be about or slightly below 2.
 \citep{Hurford}.}

 If the moon was not captured by the planet but formed alongside, the parameter $A$ assumes the value $2.2$ appropriate for rubble piles, because freshly accreted layers of a moon in formation are rubble.
So we agree to use the expression
\ba
 r_R\,=\;
 2.2\;R_m\left(\frac{M_p}{M_m}\right)^{1/3}\,\;.
 \label{1}
 \ea

 \section{The synchronous value of the semi-major axis}

 Defined by the synchronicity condition $\,{{\bf\dot{\theta\,}}}_p=n_m\,$, the synchronous value $a^{(s)}_m$ of the moon's semi-major axis is
 \ba
 a^{(s)}_m\,=\;\left( \frac{G(M_p+M_m)}{ 
 {\stackrel{\bf\centerdot}{\theta\s}}_p^{{\rm{ 2}}}
 } \right)^{1/3}
 \;\;,
 \label{2}
 \ea
 where $\,G= 6.6743\times 10^{-11}$ m$^3$ kg$^{-1}$ s$^{-2}\,$ is Newton's gravitational constant, while $\theta_p$ and ${\bf\dot{\textstyle\theta\s}}_p
 $ are the sidereal rotation angle and rotation rate of the planet. Commonly, $a^{(s)}_m$ is referred to as the {{synchronous radius}}.

 Bear in mind that this radius is defined only for moons in orbits prograde relative to the planet's spin.
 {For a retrograde moon, the semi-diurnal tidal bulge it creates on the planet will always be pulling the moon back in its orbital motion, and thus will be working to bring it down.}\,\footnote{
 {The principal input into ${\stackrel{\bf\centerdot}{a\s}}_m$ contributed by the lunar tides in the planet is of the same sign as $\,{\stackrel{\bf\centerdot}{\theta\s}}_p\s-\s n_m\;$ \citep[Section 4.2]{BoueEfroimsky}. This result, however, was derived with $n_m$ positive definite. So, for a retrograde moon, we must set $\s{\stackrel{\bf\centerdot}{\theta\s}}_p\s$ negative~---~which again will render tidal descent.}
 }

 \section{A planet synchronised by its moon}

 As we shall see below, a moon whose mass is about one percent of the planet's mass can in some situations synchronise the planet.
 {For this to happen, two inequalities must be obeyed.}

 {First, a synchronised moon's apocentre,
 $a^{(s)}_m\s(1+e_m)$, must end up below the reduced Hill radius. For small $e_m\,$, we write this simply as}
 \ba
 a^{(s)}_m\,<\,r_H^{\,\prime}\,\;.
 \label{raz}
 \ea
  {Combined with formulae (\ref{8}), (\ref{9a}), and (\ref{2}), this inequality indicates that the planet's mean motion must be about five times smaller than its rotation rate:}
 \ba
 n_p\;<\;0.198\,\s{\stackrel{\bf\centerdot}{\theta\s}}_p\,\s(1\s-\s 1.03\, e_p)^{3/2}\,\;.
 \label{inequality.eq}
 \ea

 Second, a synchronised moon's pericentre, $a^{(s)}_m\s(1-e_m)$, must be higher than the planet's Roche radius $r_R$. For low $e_m\,$, it is sufficient to simply compare $a^{(s)}_m$ with $r_R$. By inserting Eqs. (\ref{1}) and (\ref{2}) into the inequality
 \ba
 r_R\,<\,a^{(s)}_m
 \label{borderline}
 \ea
 and approximating the moon's density with that of our own Moon, we find (see Appendix \ref{Appendix A}) that the planet's rotation rate
 {in the end state, when it is synchronised by the moon,} must obey
 \ba
 {\stackrel{\bf\centerdot}{\theta\s}}_p \,<\, 1.07\;\,\mbox{hr}^{-1}\;\,.
 \label{given}
 \ea
 This sets a lower bound on the length of the sidereal day of the planet:
 \ba
 T_p\,=\,\frac{2\,\pi}{\; {\stackrel{\bf\centerdot}{\theta\s}}_p }\,>\,5.87\;\mbox{hr}\;\,.
 \label{day.eq}
 \ea
 The sidereal year of almost all known exoplanets satisfies the same restriction: \footnote{~As of November 2021, of the 4409 exoplanets with orbital periods $P_p$ determined, 4405 have periods longer than $5.87$ hr. \,Exceptions from this rule are: K2-137~b, \,KIC~10001893~b, \,KOI-55~b, \,PSR J1719-1438~b.}
 \ba
 P_p\,=\,\frac{2\,\pi}{ \,n_p }\,>\,5.87\;\mbox{hr}\;\,.
 \label{year.eq}
 \ea
   Bear in mind that we are addressing the planets synchronised by their moons, not stars. So the coincidence of the low bounds (\ref{day.eq}) and (\ref{year.eq}) in no way implies the equality of ${\stackrel{\bf\centerdot}{\theta\s}}_p\s$ and $n_p\s$.

 A close-in planet
 lacking a moon or having a moon incapable of synchronising the planet either will be tidally synchronised by the star or will end up in a higher spin-orbit resonance, if it has a sufficient eccentricity \citep{2018ApJ...857..142M}.
 According to Eq. (150) from \citet{BoueEfroimsky}, the tides in a synchronised planet will be working to reduce the planet's semimajor axis. The planet will therefore spiral in and get engulfed by the star.

 As an interesting aside, lower bound (\ref{year.eq}) on the planet period is remarkably close to the shortest rotation periods of field solar-type stars \citep{2020AN....341..497S}. This suggests that a moon-less planet, if it migrates too close to the star, is not capable of synchronising the star rotation, and therefore gets destroyed. Indeed, had such a planet been able to orbit faster than $5.87$ hr and to synchronise the star, then stars with such short rotation periods would have been observed~---~which is not the case.

 On the other hand, a planet with a period of $\sim 6$ h can hardly harbour a moon, because the gap between the planet's radius and reduced Hill radius (Sect. \ref{hill.sec}) vanishes in most cases or becomes quite small. We estimate that for our working sample of 580 inner exoplanets, the reduced Hill radius computed for the critical value of orbital period is above the actual planet's radius for just 26 systems, with only 5 of them having a niche for exomoons that is wider than $1\,R_p$.

 \section{The niche for exomoons\label{niche.sec}}

 For a moon to survive on an orbit around a planet, two conditions must be fulfilled.
 {They may come out stronger than (\ref{raz}) and (\ref{borderline}), because those were set on the end-state~---~while here we intend to address moons' parameters prior to synchronisation.}
 \begin{itemize}
 \vspace{2mm}
   \item [{\bf (1)}~] Its apocentre must be confined to the reduced Hill sphere:
   \ba
   a_m\;(1+e_m)\;<\;
     r^{\,'}_H\;\,;
   \label{10}
   \ea
   \item [{\bf (2)}~] Its pericentre must stay above the Roche sphere:
   \ba
     a_m\;(1-e_m)\;>\;
      r_R
   \;\,.
   \label{11}
   \ea
 \end{itemize}
 The following sequence of inequalities ensues:
 \ba
 r^{\,'}_H
 \,>\,a_m\,
 >\,r_R\,\;.
 \label{13}
 \ea

\begin{figure}
   \resizebox{\hsize}{!}
         {\includegraphics 
         {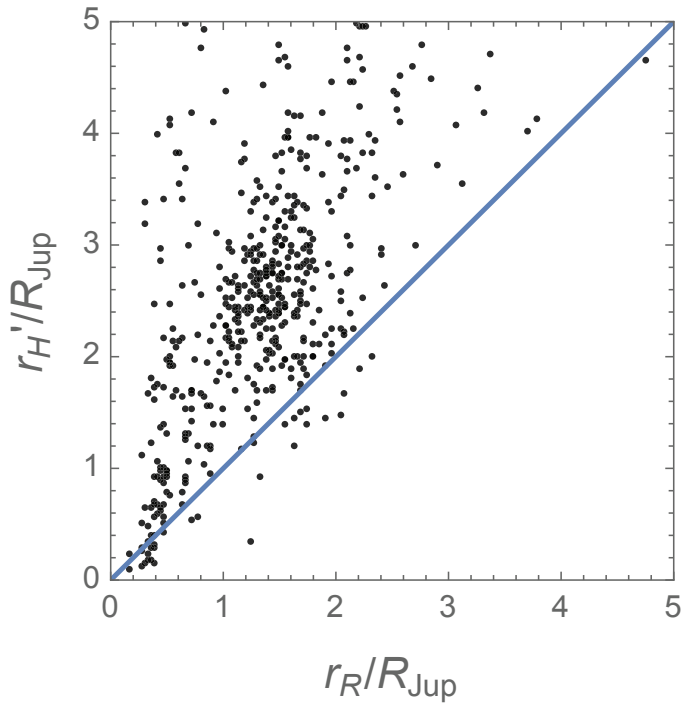}}
      \caption{Reduced Hill radii of 580 known inner exoplanets versus their estimated Roche limits in units of Jupiter's radius. The dashed diagonal line marks the locus of equal radii where the dynamical niche for satellites vanishes. The planets that fall below this line cannot have any stable satellites.}
         \label{rhrr.fig}
   \end{figure}

 Most of the known exoplanets have $r^{\,'}_H > r_R$, but not all. We find 31 planets that
 {possibly} violate this rule, and, therefore, cannot harbour any satellites. Figure  \ref{rhrr.fig} shows them as dots lying below the diagonal line of equal radii. These planets have ultra-short periods between 0.447 d (TOI-561 b) and 3.313 d (HATS-10 b), with more than half (18) being shorter than 1~d. Their eccentricities
  {are often unknown, with only upper limits given in the literature. For example, $e<0.519$ for HATS-69~b.  In our calculations involving the reduced Hill radius, this introduces additional dispersion and bias. The outlying  planets are likely to have small eccentricities below the sensitivity threshold of the transit detection method}. The planet HATS-69 b, for example, orbits an inconspicuous solar twin that does not have any known stellar companions or detectable outer planetary companions \citep{2021AJ....162..263H}, in agreement with the eccentricity migration model.
  {There are no alternative mechanisms to excite the eccentricity of close planets apart from outer massive perturbers.}
  Only one of the planets in this category, HIP 65 A b ~(with $e=0$), resides in a wide stellar binary with a smaller red dwarf companion, which apparently failed to perturb its orbit because of the significant separation $>\,230$ AU.

{Additional useful constraints on the orbital parameters of the planet-moon system can be obtained from the brackets for the moon separation. By Kepler's third law for the planet-moon and star-planet pairs, we have:}
\ba
\frac{\s M_p \, P_m^2\s}{\s M_* \,P_p^2\s } \,\simeq\, \frac{\s a_m^3\s}{a_p^3}\;.
\label{}
\ea
Combining this approximate equality with inequality (\ref{inequality}) and with the expression (\ref{8}) for the reduced Hill radius, we obtain
\ba
\frac{P_m}{P_p}< 0.198\,(1-1.03\,e_p)^{3/2}\,\;.
\label{pmpp.eq}
\ea
  {This result is stronger than formula (\ref{inequality}), because it pertains to a prograde moon at any stage of its evolution, not only at the synchronous end state (if that state is achieved).}

{Thus, prograde moons cannot have longer orbital periods than about $\s 0.2\,P_p\s$. This condition is obviously satisfied for the Earth-Moon system. We note that this upper bound on the orbital period ratio is independent of the three masses.}

{On the other hand, utilising the lower bound on the moon separation $a_m>r_R\s$, Eq. (\ref{1}), and Kepler's third law, we derive the lower bound on the orbital period of the moon in its survival niche:}
\ba
P_m\,> \,\frac{5.55}{\sqrt{\s G\,\rho_m\,}}=\left[3.26\;{\rm hr}\right]\,\sqrt{\s\rho_{\rm Moon}\s/\s\rho_m\s}\;\,,
\label{pmoon.eq}
\ea
{where $\rho_m$ is the average density of the moon and $\rho_{\rm Moon}$ is the average density of the Moon. This constraint depends only on the average density.}

 \section{Mutual synchronisation of a planet with its moon\label{above}}

 In this work, we consider the case of a planet mutually synchronised with its moon. This configuration can be attained within two different scenarios, the first of which can be initiated by various means.

 \begin{itemize}
 \item[]
 {\it Scenario 1}.~\\
 Mutual synchronism is achieved through mutual receding accompanied with a slow-down of both bodies' rotation. For this process to begin, the moon should find itself above the synchronicity orbit of the planet, and should not have too large an eccentricity.\,\footnote{~The limitation on eccentricities is needed here, because a synchronised moon with an appreciable eccentricity can move inside the synchronicity orbit, if the planet-produced tides in it are able to overpower the moon-produced tides in the planet. This is probably how Phobos crossed the synchronous orbit of Mars (\citeauthor{Bagheri} \citeyear{Bagheri}).} This initial state can be prepared by three different methods:
 \vspace{2mm}
 \begin{itemize}
 \item[(1.1)~] the moon is formed above the synchronicity orbit of the planet;
\vspace{1mm}

 \item[(1.2)~] the moon is captured above the synchronicity orbit of the planet;
\vspace{1mm}

 \item[(1.3)~] a rapidly rotating moon is formed or captured slightly below synchronism. If the initial spin is in the vicinity of the quality function peak above the synchronous rate and the separation from the synchronous orbit is small, the tides in the moon can overpower those in the planet and can push the moon above the synchronous radius.
 \end{itemize}
 \vspace{2mm}
 Whichever of the three mechanisms places the moon above the planet's synchronous orbit,
  the moon begins receding away, and the planet begins slowing down its spin.



\vspace{2mm}

 \item[]
 {\it Scenario 2}.~\\
 Mutual synchronism is achieved for a moon in a prograde orbit below the synchronous radius, or a moon in a retrograde orbit. In both cases, the tides in the planet work to shrink the moon's orbit. The tides in the moon, if its spin is synchronised with the orbital motion, work in the same direction when $e_m>0$. As the moon descends onto the planet, the planet's spin is either accelerated, for a prograde moon, or decelerated, for a retrograde moon. In the latter case, the planet may stop rotating in the sidereal frame~---~and then start rotating in the opposite direction. If the planet's angular acceleration rate is higher than the rate of moon's orbit decay, the planet's spin can be equalised with the orbital rotation before the moon reaches the Roche radius.


 \end{itemize}\vspace{2mm}

 \noindent
 Our overall objective is to understand how the star-induced tidal torque is acting on an internally synchronised planet-moon system.

 \section{Tidal torques}

 To simplify the terminology, here we often employ the adjectives `solar' and `lunar' in application to the properties of the star and the moon. For example, instead of the clumsy `star-generated tidal torque' or `moon-generated tidal torque', we say `solar torque' or `lunar torque' (with the word `tidal' omitted, because no other torques are to be included).

 \subsection{Setting}

 Our study will address a two-dimensional configuration: the lunar orbit coincides with the ecliptic, while both the planetary and lunar obliquities onto the equator of the planet are zero.
 This will enable us to consider only the polar components of tidal torques.

 A justification for the neglect of the lunar orbit inclination on the planet's orbit comes from the assumption that a significant initial inclination would make the hierarchical triple  star-planet-moon system subject to strong Lidov-Kozai interactions, which would increase interchangeably the inclinations and eccentricity of the inner pair \citep{1997AJ....113.1915I,2000ApJ...535..385F,2010ApJ...715..803V, 2011A&A...533A...7B, 2011CeMDA.111..105C, Veras2018}. As discussed in this paper, the dynamical niche for satellites is quite narrow for most known exoplanets. A large increase in eccentricity would push the reduced Hill radius limit down, possibly eliminating the niche altogether per Eq. (\ref{8}). On the other hand, close-in moons can be disrupted if the pericentre distance becomes smaller than the Roche radius. This factor further diminishes the expected frequency of exomoons. The surviving moons are
   {thus} likely to have orbits nearly coplanar with the ecliptic.

 Depending on the balance of the solar and lunar torques acting on the close planet, their  latitudinal components should relatively quickly align the planet's rotation axis with either the planet-star or the planet-moon orbit axis (see Sect. \ref{2path.sec}). In the latter case, the moons
 {may}
 be long-term viable near close planets. A limited nodal precession and nutation of the planet caused by the residual latitudinal component of the tidal torque created by the star adds to tidal dissipation in the planet and to the ensuing orbital evolution. Complex three-dimensional dynamical interactions in such nearly coplanar three-body systems are best investigated by numerical simulations, which are outside the scope of this study.

 \subsection{Secular part of the polar torque\label{3.1}}
 ~\\
 A derivation of the secular polar tidal torque from the \citeauthor{darwin1880} (\citeyear{darwin1880}) and \citet{Kaula} theory is provided in \citeauthor{Efroimsky2012} (\citeyear{Efroimsky2012}, Eq. 109), see also \citeauthor{BoueEfroimsky} (\citeyear{BoueEfroimsky}, Eqs. 123, 132, and 191):
 \ba
 \nonumber
 {\cal T}&=&2\;\frac{\s G\,{M^{\,'}}^{\,2}}{a}\;
 \sum_{{\it{l}}=2}^{\infty}~
 \left(\frac{R_p}{a}\right)^{{{2{{l}}\,+\,1}}}
 \sum_{m=0}^{ l}
 \frac{({\it{l}}\,-\,m)!}{({\it{l}}\,+\,m)!}\;m\\
  \label{31}
 \label{T31}
 \label{gedo}
 ~\\
 &\,& \sum_{p=0}^{ l}F^{\textstyle{^{\,2}}}_{lmp}(i)\sum^{\it \infty}_{q=-\infty}
 G^{\textstyle{^{\,2}}}_{{\it{l}}pq}(e)\;K_l(\omega_{lmpq})~~,~\quad~\quad
\nonumber
 \ea
 where $M^{\,'}$ is the mass of the perturber, $G$ is the Newton gravity constant, while $R_p$ is the radius of the planet.  The expression involves the inclination functions $\,F_{lmp}(i)\,$ and the eccentricity functions related to the Hansen coefficients by $\,G_{lpq}(e)\,=\,X_{l-p+q}^{\,-(l+1),~l-2p}(e)\,$,
  {see e.g., \citet[p. 232]{MurrayDermott}}.

 The quantities
 \ba
 \nonumber
 \omega_{lmpq} & \equiv & (l-2p)\dot{\omega}\s+\s(l-2p+q)n\s+\s m\s(\dot{\Omega}\s-\s\dot{\theta}_p\s)\\
 \label{omega}\\
 & \approx&  (l-2p+q)n\s-\s m\s\dot{\theta}_p\qquad
 \nonumber
 \ea
 are the Fourier tidal modes over which both the potential of the perturber and the additional tidal potential of the body are expanded. These modes' absolute values,
 \ba \chi_{lmpq}\,=\mid\omega_{lmpq}\mid\;\,,
 \label{chi}
 \ea
 are the actual physical forcing frequencies excited in the planet by the perturber.\,\footnote{~This can be observed from Eqs. (15 - 16) in \citeauthor{EfroimskyMakarov2013} (\citeyear{EfroimskyMakarov2013}).}

 The order-$l$ {{quality functions}} of the perturbed body are
 \bs
 \ba
 K_l(\omega_{lmpq})\,\equiv\,k_l(\omega_{lmpq})\;\sin\epsilon_l(\omega_{lmpq})\,\;,
 \label{notation}
 \ea
 where both the Love numbers $\s k_l\s$ and the phase lags $\s\epsilon_l\s$ are functions of the Fourier tidal modes $\s\omega_{lmpq}\s$. They can also be written as
 \footnote{~Be mindful that \,Sign$\s\epsilon_l(\omega_{lmpq})\s=\s$ Sign$\s\omega_{lmpq}\,$, \,see Eq. (24) in \citeauthor{EfroimskyMakarov2013} (\citeyear{EfroimskyMakarov2013}).}
 \ba
 K_l(\omega_{lmpq})\,\equiv\,\frac{k_l(\omega_{lmpq})}{Q_l(\omega_{lmpq})}\,\operatorname{Sign}\s(\omega_{lmpq})\,\;,
 \label{where}
 \ea
 \es
 where $Q_l$ are the tidal quality factors introduced via
 \ba
 Q_l^{-1}(\omega_{lmpq})\,=\,\mid\sin\epsilon_l(\omega_{lmpq})\mid\,\;.
 \label{q}
 \ea

 Phase lags $\s\epsilon_l(\omega_{lmpq})\s$ being odd functions, their absolute values $\s Q_l^{-1}(\omega_{lmpq})\s$ are obviously even. Even are also the Love numbers
 $\s k_l(\omega_{lmpq})\s$. So, altogether, the quality functions are odd. A function $K_l(\omega_{lmpq})$ changes its sign when the system is crossing the $lmpq$ spin-orbit resonance, one defined by $\omega_{lmpq}=0\s$. On this crossing, the passage of the function $K_l(\omega_{lmpq})$ through the zero value goes rapidly but smoothly (see Eq. \ref{smooth}).

 Since the functions $\s k_l(\omega_{lmpq})\s$ and $\s Q_l^{-1}(\omega_{lmpq})\s$ are even, we always can regard both the Love numbers and quality factors as functions of the positive definite physical frequencies (\ref{chi}):
 \ba
 k_l(\omega_{lmpq})\s=\,k_l(\chi_{lmpq})~~~,\quad Q_l(\omega_{lmpq})\s=\,Q_l(\chi_{lmpq})~\,.
 \label{}
 \ea
 Combined with Eq. (\ref{where}), this gives:
 \ba
 K_l(\omega_{lmpq})\s=\s K_l(\chi_{lmpq})\,\operatorname{Sign}(\omega_{lmpq})\,\;.
 \label{when}
 \ea

\begin{figure}
   \resizebox{\hsize}{!}
         {\includegraphics 
         {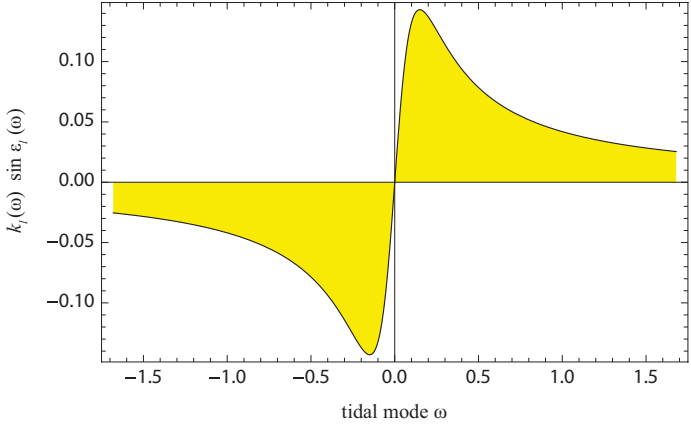}}
      \caption{Typical shape of the quality function $\s K_l(\omega)\s=\s k_l(\omega)\,\sin\epsilon_l(\omega)\s$, \,where $\s\omega\s$ is a shortened notation for the tidal Fourier mode $\s\omega_{\textstyle{_{lmpq}}}\s$. ~~(Reprinted from Icarus, Vol. 241, Noyelles et al., Spin-orbit evolution of Mercury revisited, Pages 26-44, Copyright (2014), with permission from Elsevier.)}
         \label{figure}
   \end{figure}

 The quality functions appear not only in the expression for the tidal torque, but also in the formulae for the tidal heating rate and the tidal evolution rates of orbital elements.
 A detailed discussion of the generic shape of a quality function $K_l(\omega_{lmpq})$ is provided in Appendix \ref{Appendix E}. There, it is explained that such a function has the form of a kink, as in Fig. \ref{figure}. For simple rheologies (Maxwell, Andrade), the kink has only one sharp maximum on the right and one sharp minimum on the left. These {{principal extrema}} come into being due to interplay of self-gravitation with rheological response.\,\footnote{~Gravity is working to pull the tidal bulge down, thus adding to the rigidity of the tidally perturbed body. Weak at frequencies much higher than the inverse Maxwell time, this effect becomes relevant at lower frequencies~--- and leading in the zero-frequency limit.} For more elaborate rheological models containing Debye peaks (i.e., Sundberg-Cooper or Burgers), additional maxima~---~and symmetrically located minima~---~will emerge. Additional local extrema show up also when a multi-layer structure of the body is taken into consideration.\,\footnote{~As was demonstrated by \citet{gevorgyan2021}, the tidal response of a homogeneous body that obeys the Sundberg-Cooper model is identical to the response of a body consisting of two Maxwell or Andrade layers that have different relaxation times. In application to the Moon, this correspondence was explored in detail by \citet{Walterova}.} One way or another, the principal extrema shown in Fig. \ref{figure} exist for any realistic rheology and supersede in magnitude other possible peaks.

 Dictated by hydrodynamical effects, tidal response of gas planets is much more complex. Nonetheless, qualitatively, the behaviour near a resonance must be similar: the largest peaks are defined, at large, by combination of material response and self-gravitation, and the transition through zero must be smooth.

 As demonstrated in Appendix \ref{Appendix E} for a homogeneous Maxwell sphere
  {(in realistic situations, also for a homogeneous Burgers, Andrade, or Sundberg-Cooper sphere, see Footnote \ref{footnote_models})}, the depicted in Fig. \ref{figure} two extrema of $K_l(\omega)$ are positioned at
 \ba
 {\omega_{peak}}_{\textstyle{_l}}\,=\;\pm\;\;\frac{\tau_M^{-1}}{1\,+\,{\cal{A}}_l} \,\;,
 \label{hw}
 \ea
 the corresponding extrema being
 \ba
 K_l^{\rm{(peak)}}\s=\;\pm\;\frac{3}{2\s(l-1)}\;\frac{ {\cal A}_l }{ 1 + {\cal A}_l }\,\;.
 \label{peak}
 \ea
 In these expressions, ${\cal A}_l$ are {\it dimensionless effective rigidities} rendered by Eq. (\ref{A}). Proportional to the mean shear rigidity $\mu$, these parameters serve as dimensionless measures of rheology-dominated versus gravity-dominated tidal responses, see Sects. 4.1 and 5.2 in \citet{Efroimsky2015}. For the Earth, Mars, and the Moon, the dimensionless parameter $\s{\cal A}_2\s$ assumes the values of $\s2.2\s$, $\,19\s$, and $\s64.5\s$, correspondingly.  Assuming that an exomoon is smaller than Mars, we have~\footnote{~Approximations (\ref{foramoon}) and (\ref{formoons}) may be not so exact for middle-sized and large planets. For example, a hypothetical super-Earth of the same rheology as the Earth and of a twice larger radius, will have $\s{\cal A}_2=0.55\s$ and, consequently, $\,{ {\cal A}_l }/{ (1 + {\cal A}_l) }\simeq 1/3\,$ and $\,K_l^{(peak)}\s\simeq\,\pm\,\frac{\textstyle 1}{\textstyle 4\s(l-1)}\,$.}
 \ba
 \frac{ {\cal A}_l }{ (1 + {\cal A}_l) }\simeq 1~~~,~~~\mbox{for moons}\,\;,
 \label{foramoon}
 \ea
 {
 wherefrom
  \ba
 K_l^{(peak)}\s\simeq\,\pm\,\frac{\textstyle 3}{\textstyle 4\s(l-1)}~~~,~~~\mbox{for moons}\;\,,
 \label{formoons}
 \ea
 an observation to be utilised down the road.
 }

 Equation (\ref{peak}) also renders an overall limit on the values of a quality function:
 \ba
 |K_l|\s<\s\frac{\textstyle 3}{\textstyle 4\s(l-1)}\,\;.
 \label{overall}
 \ea
 Within the narrow inter-peak interval, the functions $K_l(\omega)$ are near-linear:
 \ba
 \label{smooth}
 |\s\omega\s|<|\s{\omega_{peak}}_l\s|\,\;\Longrightarrow\,\; K_l(\omega)\s
 \simeq\s\frac{3}{2\s(l-1)}\;\frac{ {\cal A}_l }{ 1 + {\cal A}_l }\;\frac{\omega}{|\s{\omega_{peak}}_l\s|}\;,
 \ea
 and fall off as $\omega^{-1}$ outside it:
 \ba
 |\s\omega\s|>|\s{\omega_{peak}}_l\s|\,\;\Longrightarrow\,\;
 K_l(\omega)
 \s\simeq\s\frac{3}{2\s (l-1)}\;\frac{ {\cal A}_l }{ 1 + {\cal A}_l }
 \;\frac{|\s{\omega_{peak}}_l\s|}{\omega}\;\;,
 \label{od}
 \ea
 an expression to be employed shortly.

 \subsection{
 Approximation valid under no synchronism\label{3.2}}
~\\
 The secular part of the polar torque may be approximated with
 \ba
 \nonumber
 {\cal T}&=&{\cal T}_{\textstyle{_{\textstyle_{\textstyle{_{l=2}}}}}}
 \,+~O\left(\,\epsilon\,(R/a)^{7}\,\right)~\\
 \label{}\\
 &=&{\cal T}_{\textstyle{_{\textstyle{_{(lmp)=(220)}}}}}\s
   +\;\,O(\epsilon\,i^{\, 2}) \;+\; O\left(\,\epsilon\s(R/a)^7\s\right)~~,\qquad
 \nonumber
 \ea
 where $\epsilon$ is a typical value of a phase lag (\citeauthor{Efroimsky2012} \citeyear{Efroimsky2012}, Eq. 114). 
 Details on this are provided in Appendix \ref{Appendix C}, both for the solar torque acting on the planet and for the lunar torque acting on the planet. When the planet is not synchronised with either orbital motion, the leading parts of the lunar and solar torques are
 \ba
 \label{lunar 1}
 {\cal T}^{\;^{(m)}}_{\textstyle{_{\textstyle_{\textstyle{_{(lmp)=(2200)}}}}}}
 =\frac{3}{2}~\frac{G\,M_m^{\,2}}{a_m}\,\left(\frac{R_p}{a_m}\right)^5\,
 K_2(\s 2\s n_m\s-\s 2\s{\stackrel{\bf\centerdot}{\theta\s}}_p\s )
 \label{m}
 \ea
 and
 \ba
 \label{stellar 1}
 {\cal T}^{\;(\ast)}_{\textstyle{_{\textstyle_{\textstyle{_{(lmp)=(2200)}}}}}}
 =\frac{3}{2}~\frac{G\,M_{\ast}^{\,2}}{a_p}\,\left(\frac{R_p}{a_p}\right)^5\,
 K_2(\s 2\s n_p\s-\s2\s{\stackrel{\bf\centerdot}{\theta\s}}_p\s )\;\;,
 \label{star}
 \ea
 correspondingly. Hence, in neglect of $O(e^2)$, the planetary rotation is described by
 \ba
 \nonumber
 {\stackrel{\bf\centerdot\centerdot}{\theta\s}}_p & = & \frac{3}{2}\;G\;\xi^{-1}\,M_p^{\s-1}\,R_p^{\s 3}\,\left[\,M_m^{\s 2}\,a_m^{-6}\,K_2(\s 2\s n_m-\s 2\s {\stackrel{\bf\centerdot}{\theta\s}}_p\s)
 \right.
 ~\\
 \label{expression_1}\\
& \, & \qquad\,\qquad\,\qquad
\left. +\;\, M_*^{\s 2}\,a_p^{-6}\,K_2(\s 2\s n_p-\s 2\s {\stackrel{\bf\centerdot}{\theta\s}}_p \s)
 \,\right]\;\,,
 \nonumber
 \ea
 $\xi$ being the prefactor of the moment of inertia $\xi\s M_p\s R_p^{\s 2}$ of the planet. The lunar and stellar inputs in expression (\ref{expression_1}) relate as
 \ba
 \frac{\;\,{\cal T}^{\;^{(m)}}_{\textstyle{_{\textstyle_{\textstyle{_{(lmp)=(2200)}}}}}}\;\,
 }{
 {\cal T}^{\;(\ast)}_{\textstyle{_{\textstyle_{\textstyle{_{(lmp)=(2200)}}}}}}
 }\;=\;
 \left(\s\frac{M_m}{M_{\ast}}\s\right)^2\,\left(\s\frac{a_p}{a_m}\s\right)^6\,\frac{K_2(\s 2\s n_m-\s 2\s {\stackrel{\bf\centerdot}{\theta\s}}_p\s)}{K_2(\s 2\s n_p-\s 2\s {\stackrel{\bf\centerdot}{\theta\s}}_p\s)}\;\,.
 \label{numerator}
 \ea
 If the moon is massive and close enough, it can synchronise the planet. As the system is approaching synchronicity (i.e.  $\,{\stackrel{\bf\centerdot}{\theta\s}}_p\s\rightarrow\s n_m\,$), the quality function $\,K_2(\s 2\s n_m-\s 2\s {\stackrel{\bf\centerdot}{\theta\s}}_p\s)\,$ in the numerator of (\ref{numerator}) goes through a sharp peak and rapidly approaches zero (see Fig. \ref{figure}). The solar torque will then try to drive the planet out of its synchronism with the moon. Whether it will succeed in doing this will depend on the ratio between the {{peak value}} of the lunar torque and a {{non-peak}} value of the solar torque.

 \section{Two pathways to a synchronous planet--moon system\label{2path.sec}}

 As we explained in Sect. \ref{above},  mutual synchronism of the planet and the moon can be attained within two distinct dynamical scenarios having different initial conditions.

  \subsection{Scenario 1: The moon is initially on a wide prograde orbit}

  The first pathway to the 1:1 inner resonance is for the moons that are initially above the synchronous radius. By construction, the initial mean motion of such a moon falls short of the initial rotation rate of the planet: $\, n_m\,<\, {\stackrel{\bf\centerdot}{\theta\s}}_p\;$.
  \,Both the tides raised in the planet by the moon and the tides raised in the planet by the star are working in the same direction~---~to slow down the spin of the planet and, consequently, to increase its synchronicity radius. The moon is receding from the planet (and, possibly, is increasing its eccentricity $e_m\s$). The planet-moon 1:1 spin-orbit resonance can be reached if the growing synchronous radius catches up with the expanding orbit while the moon is still within the reduced Hill sphere.  We now show that the vicinity of this resonance may be stable under certain conditions, which are plausible.

\subsubsection{Approach to synchronism}

 Under the exact planet--moon synchronicity, i.e., for $\s n_m = {\stackrel{\bf\centerdot}{\theta\s}}_p\;$, \,the $e^0$ part of the lunar torque (\ref{moon1}) vanishes, while the $O(e^2)$ part of this torque survives and becomes leading. It writes as:
 \ba
 \label{Tm.eq}
 {\cal T}^{\;^{(m)}}_{\rm moon-synchr}
 =12~e_m^2~\frac{G\,M_m^{\,2}}{a_m}\,\left(\frac{R}{a_m}\right)^5\,
 K_2(\s n_m\s)\;+\;O(e_m^4\epsilon)\;\,,
 \label{torque}
 \ea
 the stellar torque still being given by expression (\ref{stellar 1}).

 Quadratic in eccentricity, torque (\ref{torque}) may appear to be much smaller than the torque exerted by the star. However, it is not the small torque at the exact resonance but the maximum torque value at a slightly shifted peak of the tidal quality function that matters in the subtle mechanism of resonance capture.

 In the vicinity of the 1:1 resonance, the quadrupole tidal quality of the planet, $K_2(\omega_{2200})$,  is an odd function of the semi-diurnal tidal mode
 \ba
 \omega_{2200}\s=\s2\,(n_m-\s{\stackrel{\bf\centerdot}{\theta\s}}_p)\,\;,
 \ea
 as in Fig. \ref{figure}, with two
 symmetric extrema bracketing the point of resonance crossing.
 Were it not for the star, the 1:1 planet-moon spin-orbit resonance would imply $\s n_m={\stackrel{\bf\centerdot}{\theta\s}}_p\,$, \,and consequently $\omega_{2200}=0\s$, \,and therefore $\s{\cal T}^{\;^{(m)}}_{\textstyle{_{\textstyle_{\textstyle{_{(lmp)=(2200)}}}}}}=\s 0\s$ according to Eq. (\ref{lunar 1}).
 In a triple system under consideration, the tides generated by the star in the planet are working to drive the planet out of its spin-orbit resonance with the moon. Aiming to synchronise the planet not with the moon but with the star, the solar tides are slowing down the spin of the planet. For this reason, the tidal mode $\omega_{2200}$ is not exactly zero but acquires a slightly positive value. The value of the quality function $K_2$ begins to climb, from the left, the steep positive peak in Fig. \ref{figure}.
 Therefrom emerges a positive lunar torque compensating for the negative solar torque~---~and the equilibrium stays stable.

 For the planet to leave said equilibrium, that is, to break out of the resonance, the tidal torque from the star has to overpower the peak tidal torque from the moon. The maximal tidal quality value achieved at the peak frequency depends on the effective rigidity ${\cal A}_l$ via Eq. (\ref{peak}). For $l=2$,
   {the tidal quality of the planet}
 cannot exceed $3/4$,
   {for a homogeneous planet}. For our purposes, it will be justified to use in the following estimates the maximal possible value of the torque exerted on the planet by the moon. Using this value, we write the ratio of the torques (given by Eqs. \ref{formoons} and \ref{stellar 1}, correspondingly) wherewith the star and the moon are acting on the planet:
 \ba
 \frac{ |\,\cal{T}^{(*)}\,|}{{\cal T}^{(m,\;{\rm peak})}}\simeq
 \frac{4}{3}\left(\s\frac{M_{\ast}}{M_m}\s\right)^2\,\left(\s\frac{a_m}{a_p}\s\right)^6\,\big{|}\s K_2(\s 2\s n_p-\s 2\s {\stackrel{\bf\centerdot}{\theta\s}}_p s)\,\big{|}\;\,.
 \label{tstm.eq}
 \ea
 {For the moon to remain locked in the 1:1 spin-orbit resonance, this ratio must be less than unity.}

\begin{figure}
   \resizebox{\hsize}{!}
         {\includegraphics 
         {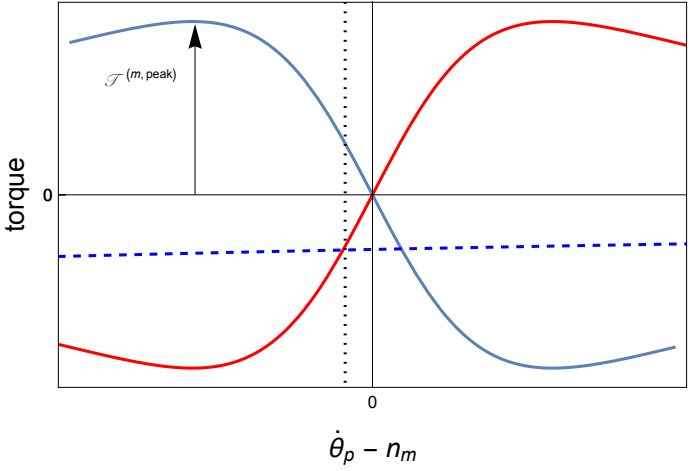}}
      \caption{Schematic depiction of the three torques involved in the quasi-stable pseudo-synchronous equilibrium of the planet-moon system. The tidal torque wherewith the moon is acting on the planet (solid blue line) is coupled with an opposite torque wherewith the planet is acting on the orbit of the moon (red curve). The negative tidal torque on the planet from the star (dashed blue line) is weakly dependent on the relative frequency. The vertical dotted line indicates the location where the tidal torques acting on the planet are equalised. Secular equilibrium is achieved between this location and 0.}
      \label{torques.fig}
   \end{figure}

 {The equilibrium state of a synchronised planet--moon system depends on the composition of the planet. To illustrate this, we resort to Fig. \ref{torques.fig}. There, the blue solid line depicts the leading, semi-diurnal component $(\ref{m})$ of the lunar tidal torque applied to the planet.
 Proportional to the semi-diurnal quality function $\s K_2(\omega_{2200}) = K_2(\,2(n_m-{\stackrel{\bf\centerdot}{\theta\s}}_p)\,)\s$, this torque naturally comes out kink-shaped. Bear in mind, though, that in Fig. \ref{torques.fig} this torque is given as a function of the difference $\s{\stackrel{\bf\centerdot}{\theta\s}}_p-\s n_m\s$, \,not of the semi-diurnal frequency $\s\omega_{2200}=\s 2\s(n_m-{\stackrel{\bf\centerdot}{\theta\s}}_p)\s$.
 }
  {The red solid line in Fig. \ref{torques.fig} is the opposite torque with which the planet is acting on the moon's orbit.}

 {Within Scenario 1, the moon starts above synchronism, so initially we have $\s{\stackrel{\bf\centerdot}{\theta\s}}_p - \s n_m\s>\s0\s$.
  As the moon is receding, and the planet is slowing down its rotation, the system is moving, in Fig. \ref{torques.fig}, from the right to the left. At the point of exact 1:1 resonance, where  $\s{\stackrel{\bf\centerdot}{\theta\s}}_p-\s n_m=0\s$, the tidal torque generated by the moon vanishes, while the negative torque applied by the star on the planet (the dashed blue line) is finite.
  For a rigid planet with a finite triaxial elongation, this negative solar torque gets compensated by a positive torque from the moon, caused by a constant tilt of the planet's longest axis with respect to the moon direction, which resembles the constant tilt of the Moon with respect to the mean Earth direction \citep{2011CeMDA.109...85R}. This way, a planet having a permanent dynamic triaxiality (a terrestrial or icy world) can be captured into an exact spin-orbit resonance with the moon, due to the presence of a restoring triaxial torque.}

 {Inner giant planets are more likely to be fluid, apart from a possible compact rigid core. If their effective viscosity is lower than a critical value \citep{2015ApJ...810...12M}, the planet gets locked in a pseudo-sychronous rotation, a long-term stable equilibrium where the relative frequency $\s{\stackrel{\bf\centerdot}{\theta\s}}_p- n_m\s$ changes with time very slowly, as we shall see now.}

  {Since no restoring triaxial torque is acting on a fluid planet at the point of exact 1:1 synchronism, evolution of this planet does not stop on arrival to that resonance. Indeed, for $\s{\stackrel{\bf\centerdot}{\theta\s}}_p -\s n_m\s=\s 0\s$ the net torque is negative due to the presence of the negative torque from the star. As a result of this, the angular acceleration $\s{\stackrel{\bf\centerdot\centerdot}{\theta\s}}_p\s$ is negative there. In Fig. \ref{torques.fig}, the system transcends the resonance and keeps moving leftwards~---~and approaches the dotted vertical line, which denotes a situation where the two torques acting on the planet are compensating each other. There, the angular acceleration of the planet becomes zero: $\s{\stackrel{\bf\centerdot\centerdot}{\theta\s}}_p\s=\s 0\s$. This state is still not the end point, and gets transcended too, because a negative torque from the planet (shown by the red curve) is acting on the lunar orbit. Impelled by this torque, the moon is now slowly descending, and the value of $n_m$ is increasing. Thus, the negative relative frequency $\s{\stackrel{\bf\centerdot}{\theta\s}}_p -\s n_m\s$ is decreasing, and the system keeps moving leftwards in the figure.}

  {The system eventually reaches an quasi-equilibrium point residing somewhere to the left of the dotted vertical line, but to the right of the peak tidal torque. The equilibrium point is where the relative angular acceleration equalises, i.e. $\,{\stackrel{\bf\centerdot\centerdot}{\theta\s}}_p\s=\s {\stackrel{\bf\centerdot}{n\s}}_m\s$. This equilibrium, however, will be slowly evolving. Indeed, in the interval between said peak and the dotted line, the net torque on the planet is positive, resulting in a spin-up. The net torque on the moon's orbit is negative, so the moon is very slowly descending back onto the planet. Therefore, in Scenario 1, the moon of a fluid planet is initially rapidly receding, and then, on arrival to equilibrium,  begins to slowly descend.}

 {In this complex pseudo-equilibrium, there is one source of energy (the orbital energy of the moon) and three sinks  (tidal dissipation in the planet, planet's rotation spin-up, and the expansion of the planet's orbit around the star). From the energy exchange aspect, the necessary condition for this  equilibrium to be stable is $\s{\dot E}_{\rm kin}\s/\s|{\dot E}_{\rm orb}|\s<\s1\s$, that is to say, the rate of the released orbital energy should be greater than the rate of the planet's rotational energy. This condition coincides with the requirement that the right-hand part of Eq. (\ref{3ksi.eq}) should be less than 1. The deviation from the exact synchronicity is expected to slowly diminish due to the decay of the moon's orbit and the corresponding increase of the amplitude of the tidal torque.}

 The relative contribution of the lunar tide (as compared to the solar one) diminishes and becomes minimal near the reduced Hill radius. Using inequality (\ref{inequality}) and the Kepler law, we
 {obtain the condition on the torque ratio required for the moon to remain locked in the 1:1 spin-orbit resonance up to the Hill radius as}
 \ba
 \frac{ |\,\cal{T}^{(*)}\,| }{{\cal T}^{(m,\;{\rm peak})}}\leq
 0.002 \s\left(\s\frac{\textstyle M_p}{\textstyle M_m}\s\right)^2\,\big{|}\s K_2(\s 2\s n_p-\s 2\s {\stackrel{\bf\centerdot}{\theta\s}}_p\s)\,\big{|}\,<\,1\,\;.  \label{tstmlimit.eq}
 \ea
  For realistic moons whose mass is about 1\% of the planet's mass, this condition is fulfilled if the tidal quality of the planet at the solar-tide frequency satisfies
 \ba
 \big{|}\s K_2(\s 2\s n_p-\s 2\s {\stackrel{\bf\centerdot}{\theta\s}}_p\s)\,\big{|}\,<\,\frac{1}{20}\,\;.
 \label{prograde}
 \ea
 Inequality (\ref{prograde}) is easily obeyed even by Earth-like terrestrial planets, because the rate of the planet's rotation in our scenario is
 \ba
 {\stackrel{\bf\centerdot}{\theta\s}}_p \simeq n_m \gg n_p\,\;,
 \ea
 and therefore the semi-diurnal tidal frequency
 \ba
 \chi\,=\,\chi_{2200}\,=\,\mid\omega_{2200}\mid\,=\,\mid 2n_p-2{\stackrel{\bf\centerdot}{\theta\s}}_p \mid\,\simeq\,2\s n_m
 \label{}
 \ea
 is much greater than the degree-two peak frequency
   \ba
   {\chi_{\rm peak}}_2\s=\s\mid{\omega_{peak}}_2  \mid\,=\,
 \;\frac{\tau_{_M}^{-1}}{1\,+\,{\cal{A}}_2}
 \,\;,
 \label{}
 \ea
 which is close to zero. For $\s\chi \gg{\chi_{\rm peak}}_2\,$, the quadrupole quality function is given by expression (\ref{od}), with $\s l=2\,$:
   \ba
   K_2(\omega)\,=\, \frac{3}{2}\; \frac{ {\cal A}_2}{1+ {\cal A}_2}\;\frac{\textstyle\mid{\omega_{peak}}_2\mid}{\textstyle\omega}\,=\,\frac{3}{2}\; \frac{ {\cal A}_2}{\left(1+ {\cal A}_2\right)^2}\;\frac{{1}}{\s\omega\s\tau_{_M}}
   \,\;.
   \ea
 The third fraction in this expression, $\,{{1}}/\left({\omega\s\tau_{_M}}\right)\,$, assumes very small values for planets with the mean Maxwell times longer than their orbital periods around the host stars (like the Earth or Mars~---~planets whose Maxwell times are in the hundreds of years).  This warrants the smallness of $K_2(\omega)$ of these planets.


 The mean Maxwell times of hot Jupiters and hot super-Earths may be comparable to or even shorter than these planets' orbital periods, in which case the factor $1/\left(\omega\s\tau_{_M}\right)$ is not warranted to be small. For such planets, however, the effective rigidities ${\cal A}_2$ are very large
  \footnote{~From Eq. (\ref{A}), we have $\,{\cal A}_l\propto\mu$. Also, within the Maxwell model an approach to viscosity dominance (i.e., to the lack of shear rigidity) implies $J\longrightarrow 0$, which is $\mu\longrightarrow\infty\;$ \citep[Section 5.2.3]{Efroimsky2015}.}. Hence, the ratio $\,{ {\cal A}_2}/\left(1+ {\cal A}_2\right)^2\,$ is vanishingly small, and the condition for resonance is still fulfilled, even at the maximum planet-moon separation equal to the reduced Hill radius.

 \subsubsection{Competition in speed\label{subsection}}

 In Scenario 1, the 1:1 resonance, $\s n_m \simeq {\stackrel{\bf\centerdot}{\theta\s}}_p\s$, must be attained before the moon reaches its reduced Hill radius. Since in this scenario the moon starts above synchronism ($\s{\stackrel{\bf\centerdot}{\theta\s}}_p - n_m > 0\s$), a configuration with $\s{\stackrel{\bf\centerdot}{\theta\s}}_p -n_m = 0\s$ can be achieved only if
 $\s\frac{\textstyle d}{\textstyle dt}\s\left({\stackrel{\bf\centerdot}{\theta\s}}_p - {n}_m\right) < 0\s$, that is to say, if
 \ba
 {\stackrel{\bf\centerdot\centerdot}{\theta\s}}_p\,<\,{{\stackrel{\bf\centerdot}{n\s}}_m}\,\;.
 \label{}
 \ea
 Both these rates being negative, we can rewrite the above as
 \ba
 |\s{{\stackrel{\bf\centerdot\centerdot}{\theta\s}}_p}\s |\,>\,|\s{{\stackrel{\bf\centerdot}{n\s}}_m}\s |\,\;.
 \label{40}
 \ea
 Thus, the absolute rate of the planet's spin-down (which is equivalent to the {\it lengthening of the day} on Earth) must be higher than the absolute rate of moon's orbital slowdown.

 As demonstrated in Appendix \ref{derivation}, in a two-body planet-moon system
 {(i.e., in neglect of the torques from star)} the ratio of the orbital expansion and the planet's spin-down rate, for $e_m=0\s$ and $\,M_p\s M_m/(M_p+\s M_m)\s\approx\s M_m\,$, \,is:
 \ba
 \frac{{\stackrel{\bf\centerdot}{n\s}}_m}{{\stackrel{\bf\centerdot\centerdot}{\theta\s}}_p}\,\approx\,3\,\xi\,\frac{M_p}{M_m}
 \,\left(\frac{R_p}{a_m}\right)^2\;\;,
 \label{3ksi.eq}
 \ea
 which is independent of the tidal quality or a specific tidal model.
 {Within the considered scenario, both the orbital angular acceleration in the numerator and the spin acceleration in the denominator of this expression are initially negative.}

 From Eq. (\ref{40}), the necessary condition for the moon's survival is that the value of ratio (\ref{3ksi.eq}) must become less than 1,
 {beginning from some value $a_m$.
 } Assuming for an estimate that the factor $3\s\xi$ equals 1,
 {we get a limit on the moon's semi-major axis value, starting from which inequality (\ref{40}) must become valid:
 \ba
 a_m>a_{\rm crit}=R_p\sqrt{\frac{M_p}{M_m}}\,\;.
 \label{rcrit.eq}
 \label{51}
 \ea
 If condition (\ref{40}) never gets obeyed, the moon will leave the reduced Hill sphere before synchronising the planet, and will eventually be lost. On the other hand, the fulfilment of condition (\ref{40}) since the birth of the moon is not necessary for synchronisation to happen. Right after accretion, the moon must be located above synchronism (as required by Scenario 1), but it may still be so close to the planet, that the right-hand side of Eq. (\ref{3ksi.eq}) is larger than unity. In this situation, while both ${\stackrel{\bf\centerdot}{n\s}}_m$ and ${\stackrel{\bf\centerdot\centerdot}{\theta\s}}_p$ are negative, the moon's orbital recession is initially going faster than the
 planet's despinning. In the course of recession, $a_m$ is growing, and, according to Eq. (\ref{3ksi.eq}) the evolution of the difference $\s{\stackrel{\bf\centerdot}{\theta\s}}_p\s-n_m\s$ may reverse. If, beginning from some value of $a_m\s$, inequality (\ref{40}) gets satisfied, the moons acquires a chance to synchronise its host planet.}

A moon of 1\% relative mass should be initially separated by more than ten planet's radii. The Moon, for example, is well outside this critical radius, and the lengthening of the day is faster than the expansion rate of the lunar orbit. So the Moon will eventually synchronise the Earth.\,\footnote{~A calculation based on the angular-momentum conservation demonstrates that under synchronism the lunar semi-major axis will assume the value of $a_m = 5.72\times 10^8$ m, which will barely fit into the reduced Hill radius $r_H^{\,\prime} = 7.09\times 10^8$~m.}
Equation (\ref{3ksi.eq}) does not take into account the contribution of the solar tidal brake, which in this scenario will make $|\s{\stackrel{\bf\centerdot\centerdot}{\theta\s}}_p\s|$ larger and push the lower bound of $a_m$ slightly lower.


 \subsubsection{
 {Scenario 1 is realisable only for small planets}
 }

\begin{figure}
   \resizebox{\hsize}{!}
         {\includegraphics 
         {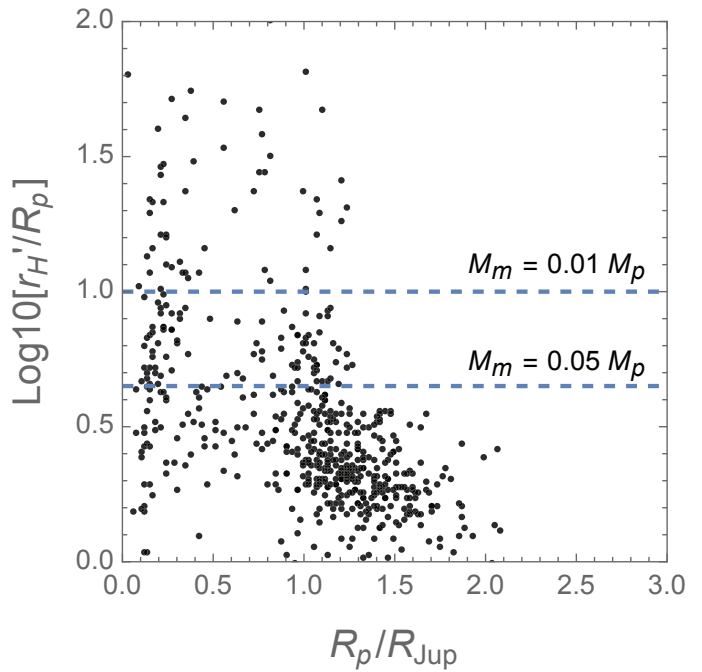}}
      \caption{Relative reduced Hill radii of 580 known inner exoplanets versus their radii in units of Jupiter's radius. Dashed horizontal lines mark the lower boundaries of plot areas where these exoplanets can in principle have moons with the smallest masses as labelled in the graph.
      {For each specific exoplanet shown with a dot in this plot, the mass of a survivable moon should be above the dashed line (above the upper line, for putative moons of $M_m=0.01\s M_p$, or above the lower line, for putative moons of $M_m=0.05\s M_p$).}}
   \label{logrh.fig}
   \end{figure}

 Replacing, in Eq. (\ref{rcrit.eq}), $a_m$ with the reduced Hill radius $r^{\,\prime}_H\s$, we estimate the minimum relative mass of the moon, $M_m/M_p$, required to achieve the critical radius within the reduced Hill sphere:
 \ba
 \frac{M_m}{M_p}\s>\s \left(\frac{R_p}{r^{\,\prime}_H}\right)^2\;\;.
 \label{52}
 \ea
 Figure
 \ref{logrh.fig} shows the results for 580 inner exoplanets from our working sample, which have all the required values in the database. The horizontal lines are shown for two specific
 $M_m/M_p$ values, 0.01 and 0.05, as benchmarks of regular and massive moons. Only those planets that are located above these lines can possibly have moons of the corresponding mass. We note
 that the majority of hot Jupiters lie below the limit even for the most imaginably massive moons,
 {because for such Jupiters inequality (\ref{52}) is violated.} Hence, they cannot be synchronised by moons,  and such systems should be long-term unstable.
 The situation is different with smaller planets. Generally, we count 58 planets that can be synchronised by a $0.01\,M_p$ moon (10\% of the general sample) and 150 (26\%) by a $0.05\,M_p$
 moon.


\subsection{Scenario 2: The moon is initially below the synchronous radius \label{2sce.sec}}

We now consider an initial configuration, in which the moon is sandwiched between the Roche radius  and the synchronous radius. The moon moves fast with respect to the planet's surface, and the tide raised by it lags the centre line. The tide brakes the orbital momentum, and the moon's orbit decays. It appears that the moon is doomed to reach the Roche limit and disintegrate. However, the same tidal interaction makes the planet's rotation accelerate. The outcome of these two parallel processes depends on the ratio of the characteristic times of orbital decay and planet's spin-up.

{In this scenario, the system from the start satisfies
 $\s
 {\stackrel{\bf\centerdot}{\theta\s}}_p\,<\,{n_m}\,$,
 where, by distinction from the previous scenario, both rates are now positive.}
 Equation (\ref{3ksi.eq}) is still valid, and this ratio must be always less than 1, in order for the moon to have a chance to synchronise the planet and survive.
  There is no possibility of survival for the moon if this ratio is initially greater than 1, because as the orbit shrinks, the ratio of the accelerations can only increase~---~so the moon will simply spiral onto the Roche limit, with the planet still rotating slower than the moon's orbital motion.
Thus, a significant initial separation between the planet and the moon (e.g., greater than $\sim 10$ planet's radii for $M_m/M_p=0.01$; cf. Eq. \ref{51}) is required. The initial separation is limited from above by the reduced Hill radius. Relatively few inner planets have sufficiently large reduced Hill radii, cf. Figures  \ref{rhrr.fig} and \ref{logrh.fig}. Thus, most of the known exoplanets, if they have moons located below the synchronous radius, would destroy these moons before their rotation could be synchronised.

The major difference between Scenario
{1 and Scenario 2 is that in the former the tidal torques from the star and from the moon, initially, are both negative, whereas in the latter, they are likely to be counter-directed. Effectively, the torque from the rapidly moving moon is initially positive in Scenario 2, and the torque from the star is negative, unless the planet initially rotates slower than its mean orbital motion.
Equation (\ref{numerator}) is valid, given that the torque from the star and the corresponding quality function are negative. The planet-moon system approaches the point of equilibrium from the left in Fig. \ref{torques.fig}, and that point is located to the left of the peak tidal torque. The general necessary condition is that the torque from the moon overpowers the torque from the star.}
The exact relation between the two torques in Scenario 2 cannot be estimated without knowing the planet's rate of rotation and its rheological properties.
If we ignore the third fraction on the right-hand side of Eq. (\ref{numerator}) for a very rough estimate, the minimum mass of the moon can be
{estimated by using condition (\ref{pmoon.eq}), assuming that the moon's average density equals that of the Moon, and requiring that the ratio of the torques should be greater than 1 in absolute value:}
\ba
 M_m>M_p\,\left[\frac{ 3.26\; {\rm hr}}{P_p}\right]^2
 \label{mmoon.eq}
 \label{54}
 \ea
 {This turns out to be a relatively loose requirement even for the close-in exoplanets. Out of 1366 planets with available estimates of $P_p$ and $M_p$ in the NASA Archive, 735 (54\%) have lower bounds on $M_m/M_{\rm Moon}$ within 1, and 834 (61\%) within 5. There is a caveat in this consideration, though. The outcome of the competition in speed also depends on the initial rate of planet's rotation ${\stackrel{\bf\centerdot}{\theta\s}}_p$ and its tidal properties. If the initial rotation is very slow, ${\stackrel{\bf\centerdot}{\theta\s}}_p\s<\s n_p$, or even retrograde,
the planet also has to cross the 1:1 spin-orbit resonance with the star before it can be synchronised by the moon. To avoid capture into this resonance, the quality function in the numerator in Eq. (\ref{numerator}), which is approximately $K_2(\s 2\s n_m-\s 2\s n_p)\s$, should not be much smaller than the peak quality of the tide caused by the star. A wide range of fairly complex scenarios emerge for planets of terrestrial composition. The most important and poorly known parameters are the effective viscosity of the mantle, how close the mantle is to solidus \citep{2018ApJ...857..142M}, and the possible contribution of the Andrade-type dissipation mechanisms.
}

{If the braking torque from the star overpowers} the torque from a small or distant moon, the planet gets synchronised by the star. The moon will then inevitably spiral in to the Roche radius and disintegrate. The characteristic times for this process strongly depends on a number of planet's parameters including its rheology. Somewhat counterintuitively, Earth-like rigid planets may be less efficient in dissipating tidal energy because of the large difference between $n_p$ and $n_m$, i.e., the high tidal frequency. The time of orbital decay for such systems (and the life time of their satellites) is also dependent on the residual eccentricity of the moon.




 \section{Discussion \label{disc.sec}}

 We have demonstrated, on very general principles, that exomoons can orbit most of known exoplanets only within a relatively thin shell of a few to several Jupiter radii (Figs. 1 and 2).

 The dynamics of triple star-planet-moon systems is determined by the interplay of direct gravitational interaction, on the one hand, and the tidally caused exchange of spin and orbital angular momenta, on the other. For various initial configurations, this interplay leads to a multitude of possible scenarios of dynamical evolution \citep{2021PASP..133i4401D,Tokadjian,Tokadjian2}. Most of them, however, predict a short-lived moon destined either to spiral in to the planet's Roche radius or to retreat from the planet and be ejected by the gravitational pull of the star. In this study, we focus on the remaining possibilities for hypothetical exomoons to survive over longer timescales. These possibilities are critically dependent on the ability of the moon to synchronise the planet, that is to say, to overpower the tidal action from the star on the planet. This scenario is not far-fetched, for it is actually epitomised by our own Moon~---~which is in the process of synchronising the Earth. The settings for most of the known exoplanets are much less accommodating, however. These planets are closer to their hosts and are larger than the Solar System planets. On the one hand, in order to exert a strong torque on the planet, a moon should be sufficiently close to it and sufficiently massive. On the other hand, a very close moon would undergo fast orbital expansion (cf. Eq. \ref{3ksi.eq}) with a likely outcome of being driven outside the Hill sphere before the planet becomes synchronised. This places an additional lower bound on the relative mass of the moon, $M_m/M_p\s$, as expressed by Eq. (\ref{52}) and depicted in Fig. \ref{logrh.fig}. For each specific exoplanet shown with a dot in this plot, the mass of a survivable moon should be well above the dashed lines
 {(above the upper line, for putative moons of $M_m=0.01\s M_p$, or above the lower line, for putative moons of $M_m=0.05\s M_p$)}. Only small fractions of the available sample satisfy this condition: 10\% at $M_m/M_p>0.01\s$, and
 26\% at $M_m/M_p>0.05$. The majority of discovered exoplanets are massive, and the median limiting mass $0.05\,M_p$ equals  $917\,M_{\rm Moon}\s$, while $0.01\,M_p$ equals to $183\,M_{\rm Moon}\s$. It is doubtful that such massive satellites are common, or that any even exist. Limiting our consideration to satellites less massive than $7\,M_{\rm Moon}\s$, we estimate that only 10\% of all known planets satisfy the $M_m/M_p>0.01$ requirement. We therefore predict that exoplanet systems stabilised by their moons should be quite rare in the current compendium, perhaps a few per hundred.

\begin{figure}
   \resizebox{\hsize}{!}
         {\includegraphics 
         {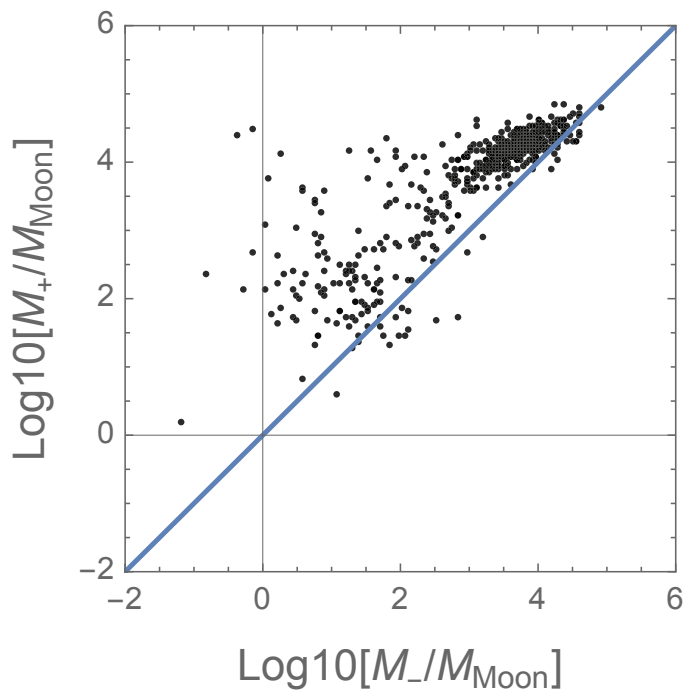}}
      \caption{Lower masses of exomoons that can in principle synchronise their host planets for 580 known exoplanet systems, in units of the mass of the Moon. The $M_-$ limits are derived from the estimated reduced Hill radii and refer to the Scenario 1 pathway, when the exomoon is initially above the synchronous radius. The $M_+$ limits are derived from the estimated Roche radii and refer to the Scenario 2 pathway, when the exomoon is initially below the synchronous radius.}
   \label{plus.fig}
   \end{figure}

 Another set of constraints on a planet-stabilising moon can be derived from the condition of relative angular acceleration (given by Eq. \ref{3ksi.eq}) being less than $1$. Irrespective of the rheological properties of the planet, a moon has a chance to synchronise its host planet only if $\dot n_m < \ddot{\theta\s}_p$. Otherwise, because of the narrowness of the exomoon niche, it is likely to bump into either the reduced Hill limit or the Roche limit, depending on the initial $\ddot{\theta\s}_p$. The initial $a_m$ is arbitrary in this equation, but it should be between $r_R$ and $r^{\,\prime}_H$. Substituting $a_m$ with these limits, and assuming $3\s\xi=1$, we can estimate the corresponding limits $M_-$ and $M_+$ for the mass of the moon. The results for 580 exoplanets with sufficient data in the database are presented in Fig. \ref{plus.fig}. We note that both these limits are one-sided, in that the mass should be greater than the limit, for the moon to survive. The $M_p>M_-$ condition is applicable in the Scenario 1 pathway, wherein the moon's orbit is initially expanding. The $M_p>M_+$ condition is applicable in the Scenario 2 pathway, wherein the moon's orbit is initially shrinking. The graph shows that the bulk of known inner exoplanets, seen as a dense cloud in the upper-right corner, require extremely massive satellites in the range of Earth-like planets for the pathways to work. We estimate the number of exoplanets with survivable moons within $5\,M_{\rm Moon}$ to be 28 (out of 580) for pathway 1, and only 2 for pathway 2. Thus, exomoons that can survive in the vicinity of known close-in planets over stellar lifetime durations should be rare.


This leads to question of how theoretical cogitations can be test using observations. Exomoons, even unusually massive ones, cannot be imaged in the glare of the close host stars. Exoplanets harbouring synchronising moons should have finite eccentricities because of the residual tidal torque couple to the star. They may also be impervious to the tidal orbital decay, which is expected
from moon-less close-in planets.
  {Depending on the balance of the tidal torques in the star-planet pair, a moon-synchronous planet may exhibit orbital expansion instead of orbital decay.} These are indirect indicators, however, which allow for an alternative interpretation. The best promise is offered by precision photometry of transiting exoplanets. The light reflected by the planet modulates the general light curve due to the planetary phases. The peak flux is expected to coincide with the upper conjunction, which may be observable as a shallow secondary transit.  
  {Hot Jupiters are also expected to irradiate substantial amounts of infrared light from their surfaces. Due to thermal inertia, rapidly rotating planets would have hot spots with phase shifts leading the planet-star directions, which would result in an observable shift between the overall light curve and the eclipse in the upper conjunction.} This, however, can be ambiguously interpreted as a finite eccentricity of the star-planet pair. Despite these difficulties, a search for exomoons has a tremendous impact on our understanding of cosmogony of other planetary worlds, and it should continue along both theoretical and practical routes. The two tidal-orbital pathways to long-term survival of exomoons in the harsh conditions of known systems suggest that the fate of close-in exoplanets may be intertwined with the fate of their satellites.


\begin{appendix}


 \section{Comparing the Roche radius and the synchronous value of the semi-major axis\label{Appendix A}}

  We wish to know for what critical spin rate ${\stackrel{\bf\centerdot}{\theta\s}}_p^{(crit)}$ the semi-major axis of the synchronous orbit of the moon about the planet,
  $a^{(s)}_m$, coincides with the planet's Roche radius $r_R$. Formulae (\ref{1}) and (\ref{2}) entail:
 \ba
 \nonumber
 r_R\,<\,a^{(s)}_m\quad
 \Longrightarrow \qquad\qquad\qquad\qquad\qquad\qquad\qquad
 ~\\
 \label{}\\
  {\stackrel{\bf\centerdot}{\theta\s}}_p\,<\,\sqrt{ \frac{4}{3}\,\pi\,\frac{\rho_m\,G}{A^3}\,}
 \,\sqrt{\,\frac{M_p+M_m}{M_p}\,}\approx\sqrt{ \frac{4}{3}\,\pi\,\frac{\rho_m\,G}{A^3}\,}\;\,.\;\;\;
 \nonumber
 \ea
 For $\,A=2.2\,$, and for the density $\rho_m$ of the moon given in kg/m$^{3}\,$, this renders:
 \ba
 {\stackrel{\bf\centerdot}{\theta\s}}_p\s <\s 5.12\times 10^{-6}\s\sqrt{\rho_m\,}\;\;\mbox{s}^{-1}=\,
 1.84\times 10^{-2}\,\sqrt{\rho_m\,}\;\;\mbox{hr}^{-1}\;~,\;\;
 \ea
 For the density of our Moon, $\rho_m=3340$~kg~m$^{-3}$, we have
 \ba
 {\stackrel{\boldmath\centerdot}{\theta\s}}_p \,<\, 1.07\;\,\mbox{hr}^{-1}\;\,.
 \label{value}
 \ea

 \section{Quality functions. A typical shape\label{Appendix E}}

 To get an idea of a typical shape of a quality function, consider a tidally perturbed homogeneous body whose rheology is defined by a complex compliance
 \ba
 \bar{J}(\chi)\,=\,{\cal R}{\it e}\left[\bar{J}(\chi)\right]\,+\,i\;{\cal I}{\it m}\left[\bar{J}(\chi)\right]\,\;,
 \label{}
 \ea
 with
 \ba
 \chi\,=\,\chi_{lmpq}
 \label{}
 \ea
 being a concise notation for a tidal frequency $\chi_{lmpq}=|\omega_{lmpq}|$ exerted by tides.

 For example, the Maxwell rheology looks like
 \ba
 ^{(Maxwell)}\bar{J}(\chi)\s=\s J\s+\s i \s\left(\s-\s\frac{1}{\eta\s\chi}  \right)\,=\s J\s\left(1\s-\s\frac{1}{\tau_{_M}\s\chi}  \right)\,\;,
 \label{Maxwell}
 \ea
 where $\eta$ is the shear viscosity, $J$ is the unrelaxed sheer compliance (the inverse of the unrelaxed shear rigidity $\mu$), and $\tau_{_M}=\eta\s J=\eta/\mu$ is the Maxwell time. In the more realistic Andrade model, one has to include terms responsible for transient processes. The elastic, instantaneous, part of the complex compliance will, though, still be $J$. Neither the Maxwell nor the Andrade model are capable of taking the relaxation of the elastic part into consideration. (This relaxation is, however, an actual effect taking place mainly due to grain-boundary sliding.) So these models fail to distinguish between the unrelaxed and relaxed values of $J$. To fix this flaw, a more detailed model by Sundberg and Cooper should be employed. A simple review on the topic is available in Section 3 of \citet{Walterova} and in
  Section 2 of \citet{TidalReview}.


 As demonstrated, for example, in  \citet[]{Efroimsky2015}, Eq. (40), the quality function of a homogeneous body is
 \footnote{~To obtain (\ref{63b}), we divided both the numerator and denominator of (\ref{63a}) by the squared elastic part $J$ of the complex compliance $\bar{J}$, to make both the numerator and denominator dimensionless.}
 \bs
 \ba
 \nonumber
 & K_{l}(\chi)  \equiv  {k}_l(\chi)\;\sin\epsilon_{l}(\chi)\\
 \nonumber\\
 & =  - \;\frac{\textstyle 3}{\textstyle 2(l-1)}~\frac{\textstyle {\cal{B}}_{\textstyle{_l}}\;\;{\cal I}{\it m}\left[\bar{J}(\chi)\right]
 }{\textstyle \left(\,{\cal R}{\it e}\left[\bar{J}(\chi)\right]+{\cal{B}}_{\textstyle{_l}\s}\right)^2+\left(\,{\cal I}{\it m}
 \left[\bar{J}(\chi)\right]\,\right)^2}
 \label{63a}\\
 \nonumber\\
 \nonumber\\
 & =  - \;\frac{\textstyle 3}{\textstyle 2(l-1)}~\frac{\textstyle {\cal{A}}_{\textstyle{_l}}\;\;{\cal I}{\it m}\left[\bar{J}(\chi)\;J^{-1}\right]
 }{\textstyle \left(\,{\cal R}{\it e}\left[\bar{J}(\chi)\;J^{-1}\right]+{\cal{A}}_{\textstyle{_l}\s}\right)^2+\left(\,{\cal I}{\it m}
 \left[\bar{J}(\chi)\;J^{-1}\right]\,\right)^2}\;\;,\;\qquad
 \label{63b}
 \ea
 \label{63}
 \es
 with the dimensional constants ${\cal{B}}_{l}$ and dimensionless constants ${\cal{A}}_{l}$ defined as
 \ba
 {\cal{B}}_{l}\,\equiv~\frac{\textstyle{(2\,l^{\,2}\,+\,4\,l\,+\,3)}}{\textstyle{l\,\mbox{g}\,
 \rho\,R}}~=\;\frac{\textstyle{3\;(2\,l^{\,2}\,+\,4\,l\,+\,3)}}{\textstyle{4\;l\,\pi\,
 G\,\rho^2\,R^2}}~\;,
 \label{B}
 \ea
 \ba
 {\cal{A}}_{l}\s\equiv\s{{\cal{B}}_{l}}\,{J^{-1}}=\s{\cal{B}}_{l}\,\mu\,\;,
 \label{A}
 \ea
 $G$ being Newton's gravitational constant, and $g$, $\rho$, $R$ being the surface gravity, density, and radius of the body.
 Be mindful that expression (\ref{gedo}) for the secular part of the polar tidal torque contains not $\s K_{l}(\chi_{lmpq})\s$ but the odd function $\s K_l(\omega_{lmpq})\s=\s K_l(\chi_{lmpq})\,\operatorname{Sign}(\omega_{lmpq})\s$.

 For the Maxwell or Andrade model, the function $\s K_l(\omega_{lmpq})$ has the shape of a kink, shown in Figure~\ref{figure}.\,\footnote{~It can also be shown from Eq. (45) in \citep{Efroimsky2015} that the  frequency-dependence of
 $\s\sin\epsilon_l(\omega)=Q^{-1}_l(\omega)\,\mbox{Sign}(\s\omega\s)$
 has a similar shape, with each of the two extrema located within less than a decade from a corresponding extremum of $K_l$.}
 Specifically, for the Maxwell rheology (\ref{Maxwell}), the quality functions become
 \ba
 K_l(\omega)\,=\,\frac{3}{2(l-1)}~\frac{ \tau_{_M}\s\omega\;{\cal A}_l }{1\,+\, \left(\tau_{_M}\s\omega\right)^2\,\left(1\s+\s{\cal A}_l^2  \right)    }\;\;,
 \label{function}
 \ea
 $\omega$ being a short notation for $\omega_{lmpq}\,$.

 From the above expression, we observe that for a Maxwell body~---~and, in realistic situations, for Andrade, Burgers, and Sundberg-Cooper bodies also\,\footnote{~Transient processes in these materials are pronounced mainly over seismic frequencies, the rheological response at low frequencies thus being close to Maxwell. Consequently, the peaks for an Andrade body are very similar, in both amplitude and location, to the Maxwell values given by Eq. (\ref{wh}).\label{footnote_models}}~---~the extrema of the kink  $\s K_l(\omega)\s$ are residing at
 \ba
 {\omega_{peak}}_{\textstyle{_l}}\,=\;\pm\;
 \;\frac{\tau_M^{-1}}{1\,+\,{\cal{A}}_l}
 \,\;,
 \label{wh}
 \ea
 the corresponding peak values being
 \ba
 K_l^{\rm{(peak)}}\s=\;\pm\;\frac{3}{4\s(l-1)}\;\frac{ {\cal A}_l }{ 1 + {\cal A}_l }\,\;.
 \label{see}
 \ea
 In the inter-peak interval, a quality function $\s K_l(\omega)\s$ is almost linear: \footnote{~The linearity of $\s K_l(\omega)\s$ is equivalent to the frequency-independence of the time lag: $\s\Delta t_l(\omega_{\textstyle{_{lmpq}}})\s=\s\Delta t\s$
 \citep{EfroimskyMakarov2013}. This is why the Constant Time Lag (CTL) tidal model can be used exclusively for $\s|\s\omega\s|<|\s{\omega_{peak}}_l\s|\,$, and is badly misleading outside this interval. Also, it is owing to this linearity of $K_l(\omega)$ in the inter-peak interval that the tidal torque value transcends spin-orbit resonances continuously, not in abrupt steps.}
 \ba
 \nonumber
 & \s &
 |\s\omega\s|\;<\;|\s{\omega_{peak}}_l\s|\quad\Longrightarrow\qquad\qquad\qquad
 ~\\
 \label{444}\\
 &\s&\phantom{~~~~~~~~~}\qquad K_l(\omega)\;
 \simeq\;\frac{3}{2\s(l-1)}\;\frac{ {\cal A}_l }{ 1 + {\cal A}_l }\;\frac{\omega}{|\s{\omega_{peak}}_l\s|}\,\;.
 \nonumber
 \ea
 Outside this interval, it falls off as the inverse $\omega\,$:
 \ba
 \nonumber
 & \s &
 |\s\omega\s|\;>\;|\s{\omega_{peak}}_l\s|\quad\Longrightarrow\qquad\qquad\qquad
 ~\\
 \label{do}
 \label{555}\\
 & \s& \phantom{~~~~~~~~~}\qquad K_l(\omega)\,
 \simeq\,\frac{3}{2\s (l-1)}\;\frac{ {\cal A}_l }{ 1 + {\cal A}_l }
 \;\frac{|\s{\omega_{peak}}_l\s|}{\omega}\;\,.\;
 \nonumber
 \ea
 From Eq. (\ref{see}), we observe that the peaks' amplitude is independent of the viscosity value $\s\eta\s$. On the other hand, the spread between the extrema is inversely proportional to $\s\eta\s$, as follows from expression (\ref{wh}). For solid terrestrial planets, the mean viscosity values are high, wherefore the extrema are located close to zero (i.e., to the resonance $\omega\equiv\omega_{nmpq}=0$), so the inter-peak interval of linear frequency-dependence of $K_l$ is very narrow and the peaks are sharp.

 For a Maxwell or Andrade body, both the kink function $K_l(\omega)$ and the similar to it function $\s\sin\epsilon_l(\omega)=
 Q^{-1}_l(\omega)\,\mbox{Sign}(\s\omega\s)$ have only one peak for a positive $\omega$.
 The situation will, however, change if we plug into formula (\ref{63}), and into its counterpart for $\,\sin\epsilon_n(\omega)=Q^{-1}_n(\omega)\,\mbox{Sign}(\s\omega\,)$, a complex compliance appropriate to a more accurate description, such as the Sundberg-Cooper rheology. In that case, an additional, smaller local maximum will appear on the right slope (an a similar local minimum will emerge on the left slope). These additional peaks may have big geophysical consequences for the Moon \citep{Walterova}.

 \section{Secular components of the solar and lunar tidal torques acting on the planet \label{Appendix B}\label{Appendix C}}

 The quadrupole part of the torque looks like (\citeauthor{Efroimsky2012} \citeyear{Efroimsky2012}):
 \ba
 \nonumber
 & \s &
 {\cal T}_{\textstyle{_{\textstyle_{\textstyle{_{(lmp)=(220)}}}}}}\,=
 ~\\
 \nonumber\\
 \nonumber\\
 \nonumber
 & \s & ~~~~~\frac{3}{2}~\frac{G\,{M^{\,\prime}}^{\,2}}{a}\,\left(\frac{R_p}{a}\right)^5\,\left[~\frac{1}{2304}~e^6~K_2 (\s-\s n\s-\s2\s\dot{\theta}_p\s)~~
 \right.\\
 \nonumber\\
 \nonumber\\
 \nonumber
 & \s & +\,\left(~ \frac{1}{4}~e^2-\,\frac{1}{16}~e^4~+\,\frac{13}{768}~e^6 ~\right)~K_2 (\s n\s-\s2\s\dot{\theta}_p\s)
 ~\\
 \nonumber\\
 \nonumber\\
 \nonumber
 & \s & + \, \left(1\,-\,5\,e^2\,+~\frac{63}{8}\;e^4\,-~\frac{155}{36}~e^6\right)~K_2 (\s 2\s n\s-\s2\s \dot{\theta}_p\s)\\
 \nonumber\\
 \nonumber\\
 \nonumber
 &\s& +\,\left(\frac{49}{4}~e^2-\;\frac{861}{16}\;e^4+~\frac{21975}{256}~e^6\right)~K_2 (\s3\s n\s-\s 2\s\dot{\theta}_p\s)\\
 \nonumber\\
 \nonumber\\
 \nonumber
 &\s& +\,\left(~ \frac{289}{4}\,e^4\,-\,\frac{1955}{6}~e^6 ~\right)~K_2(\s 2\s n\s-\s\dot{\theta}_p\s)\\
 \nonumber\\
 \nonumber\\
 &\s& + \, \left. \frac{714025}{2304}\;e^6\;K_2(\s 5\s n\s-\s3\s\dot{\theta}_p\s)
 \,\right]
 \nonumber
 ~\\
 \label{star1}
 \label{moon1}\\
 \nonumber
 &\s& +\;O(e^8\,\epsilon)\;+\;O(i^{\s 2}\,\epsilon)~~,~\quad~\quad~\quad
 \ea
 where $M^{\,\prime}$ is the mass of the perturber, $\epsilon$ is a typical value of the phase lag in the perturbed planet, and the perturber's orbital inclination $i$ on the planet's equator is small enough for the order-$i^2$ terms to be dropped.

This formula will render the solar torque ${\cal T}^{\,^{(\ast)}}$, provided we identify the mass of the perturber with that of the star: $\s M^{\,'}=M_{\,\ast}\,$, \,and employ the elements $\s a = a_p\s$, $\s e = e_p\s$, and use the mean motion $n= n_p$ of the planet's orbit.

The formula will give us the lunar torque $\s{\cal T}^{\,^{(m)}}$, if we set $\s M^{\,'}=M_{m}\s$, $\, a = a_m\s$, $\, e = e_m\s$, \,and $\s n= n_m\s$.

 \section{Derivation of Eq. (\ref{3ksi.eq})\label{derivation}}

 For the planet-moon two-body system, the angular momentum is given by
 \ba
 \nonumber
 H & = & \frac{M_p\s M_m}{M_p +\s M_m}\,\sqrt{G\s (M_p +\s M_m )\,a_m\,(1\,-\,e_m^2)\,}
 ~\\
 \label{en}\\
 & \, & +\; C_p \,{ {\stackrel{\bf\centerdot}{\theta\s}}_p }\s+\s C_m\,{\stackrel{\bf\centerdot}{\theta\s}}_m\;\;,
 \nonumber
 \ea
 $C_p$ and $C_m$ being the maximal moments of inertia of the planet and the moon, correspondingly, and ${\stackrel{\bf\centerdot}{\theta\s}}_p$ and ${\stackrel{\bf\centerdot}{\theta\s}}_m$ being their rotation rates. If we neglect the eccentricity and assume that the spin angular momentum of the moon is negligible as compared to that of the planet,
 the above expression will get simplified to
 \ba
 H\;=\;\frac{M_p\s M_m}{M_p +\s M_m}\,\sqrt{G\s (M_p +\s M_m )\,a_m\,}\s+\s C_p \,{ {\stackrel{\bf\centerdot}{\theta\s}}_p }\;\;,
 \label{route}
 \ea
 differentiation whereof will give us:
 \ba
 0\,=\,\frac{M_p \,M_m}{M_p + M_m}\;\sqrt{G\,(M_p + M_m)\,}\,\frac{d}{dt}\s a_m^{1/2}
 \,+\,C_p \,{\stackrel{\bf\centerdot\centerdot}{\theta\s}}_p \;\;.
 \label{L}
 \ea
 Casting the definition $\s n_m=\sqrt{G(M_p+M_m)\, a_m^{-3}}\,$ into the form of $a_m^{ 1/2}=(G\s(M_p+M_m))^{1/6} n_m^{-1/3}$, we derive:
 \ba
 \nonumber
 \frac{d}{dt}\s a_m^{1/2} &=&-\,\frac{1}{3}\,(G(M_p+M_m))^{1/6}\s n_m^{-4/3}\s {\stackrel{\bf\centerdot}{n\s}}_m\\
 \label{}\\
 &=&-\,\frac{1}{3}\,(G(M_p+M_m))^{-1/2} \s a_m^{2} \s{\stackrel{\bf\centerdot}{n\s}}_m\;\;,
 \nonumber
 \ea
 insertion whereof into Eq. (\ref{L}) entails
 \ba
 0\,=\,-\,\frac{1}{3}\,\frac{M_p \,M_m}{M_p + M_m}\;a_m^2\,{\stackrel{\bf\centerdot}{n\s}}_m
 \,+\,C_p \,{\stackrel{\bf\centerdot\centerdot}{\theta\s}}_p \;\;.
 \label{put}
 \ea
 Through a dimensionless coefficient $\xi$, the maximal moment of inertia is conventionally expressed as $\s C_p\s=\s\xi\s M_p\s R^2_p\s$.
 Also, since {{en route}} from Eq. (\ref{en}) to Eq. (\ref{route}) we set $\,|C_m\s{\stackrel{\bf\centerdot}{\theta\s}}_m|\ll |C_p\s{\stackrel{\bf\centerdot}{\theta\s}}_p|\s$, then in about the same approximation we may put $\,{M_p\s M_m}/(M_p + M_m)\approx M_m\s$ into Eq. (\ref{put}). This gives us
 \ba
 0\,=\,-\,\frac{1}{3}\,M_m\,a_m^2\,{\stackrel{\bf\centerdot}{n\s}}_m\,+\,\xi\s M_p\s R^2_p \,{\stackrel{\bf\centerdot\centerdot}{\theta\s}}_p \;\;.
 \label{}
 \ea

\end{appendix}

 \begin{acknowledgements}
 The authors are grateful to James L. Hilton for useful comments and recommendations.
 {We also
 thank our anonymous reviewer for a substantial and deep review, which resulted in a more consistent and focused paper}.
 \end{acknowledgements}

\bibliographystyle{aa}
\bibliography{Exomoons}

\begin{thebibliography}{39}
\expandafter\ifx\csname natexlab\endcsname\relax\def\natexlab#1{#1}\fi

\bibitem[{{Agnor} \& {Hamilton}(2006)}]{Agnor2006}
{Agnor}, C.~B. \& {Hamilton}, D.~P. 2006, Nature, 441, 192

\bibitem[{{Astakhov} {et~al.}(2003){Astakhov}, {Burbanks}, {Wiggins}, \&
  {Farrelly}}]{Astakhov}
{Astakhov}, S.~A., {Burbanks}, A.~D., {Wiggins}, S., \& {Farrelly}, D. 2003,
  Nature, 423, 264

\bibitem[{{Astakhov} \& {Farrelly}(2004)}]{AstakhovFarrelli2004}
{Astakhov}, S.~A. \& {Farrelly}, D. 2004, Monthly Notices of the Royal
  Astronomical Society, 354, 971

\bibitem[{{Bagheri} {et~al.}(2022){Bagheri}, {Efroimsky}, {Castillo-Rogez},
  {Goossens}, {Plesa}, {Rambaux}, {Rhoden}, {Walterov{\'a}}, {Khan}, \&
  {Giardini}}]{TidalReview}
{Bagheri}, A., {Efroimsky}, M., {Castillo-Rogez}, J., {et~al.} 2022, Advances
  in Geophysics, 63, 231

\bibitem[{{Bagheri} {et~al.}(2021){Bagheri}, {Khan}, {Efroimsky}, {Kruglyakov},
  \& {Giardini}}]{Bagheri}
{Bagheri}, A., {Khan}, A., {Efroimsky}, M., {Kruglyakov}, M., \& {Giardini}, D.
  2021, Nature Astronomy, 5, 539

\bibitem[{{Barnes} \& {O'Brien}(2002)}]{2002ApJ...575.1087B}
{Barnes}, J.~W. \& {O'Brien}, D.~P. 2002, The Astrophysical Journal, 575, 1087

\bibitem[{{Batygin} {et~al.}(2011){Batygin}, {Morbidelli}, \&
  {Tsiganis}}]{2011A&A...533A...7B}
{Batygin}, K., {Morbidelli}, A., \& {Tsiganis}, K. 2011, Astronomy and
  Astrophysics, 533, A7

\bibitem[{{Bou{\'e}} \& {Efroimsky}(2019)}]{BoueEfroimsky}
{Bou{\'e}}, G. \& {Efroimsky}, M. 2019, Celestial Mechanics and Dynamical
  Astronomy, 131, 30

\bibitem[{{Chandrasekhar}(1987)}]{Chandrasekhar}
{Chandrasekhar}, S. 1987, {Ellipsoidal figures of equilibrium}

\bibitem[{{Correia} {et~al.}(2011){Correia}, {Laskar}, {Farago}, \&
  {Bou{\'e}}}]{2011CeMDA.111..105C}
{Correia}, A. C.~M., {Laskar}, J., {Farago}, F., \& {Bou{\'e}}, G. 2011,
  Celestial Mechanics and Dynamical Astronomy, 111, 105

\bibitem[{Darwin(1880)}]{darwin1880}
Darwin, G.~H. 1880, Philosophical Transactions of the Royal Society of London,
  171, 713

\bibitem[{{Dobos} {et~al.}(2021){Dobos}, {Charnoz}, {P{\'a}l}, {Roque-Bernard},
  \& {Szab{\'o}}}]{2021PASP..133i4401D}
{Dobos}, V., {Charnoz}, S., {P{\'a}l}, A., {Roque-Bernard}, A., \& {Szab{\'o}},
  G.~M. 2021, Publications of the Astronomical Society of the Pacific, 133,
  094401

\bibitem[{{Domingos} {et~al.}(2006){Domingos}, {Winter}, \&
  {Yokoyama}}]{Domingos}
{Domingos}, R.~C., {Winter}, O.~C., \& {Yokoyama}, T. 2006, Monthly Notices of
  the Royal Astronomical Society, 373, 1227

\bibitem[{{Efroimsky}(2012)}]{Efroimsky2012}
{Efroimsky}, M. 2012, Celestial Mechanics and Dynamical Astronomy, 112, 283

\bibitem[{{Efroimsky}(2015)}]{Efroimsky2015}
{Efroimsky}, M. 2015, The Astronomical Journal, 150, 98

\bibitem[{{Efroimsky} \& {Makarov}(2013)}]{EfroimskyMakarov2013}
{Efroimsky}, M. \& {Makarov}, V.~V. 2013, The Astrophysical Journal, 764, 26

\bibitem[{{Efroimsky} \& {Makarov}(2022)}]{2022Univ....8..211E}
{Efroimsky}, M. \& {Makarov}, V.~V. 2022, Universe, 8, 211

\bibitem[{{Ford} {et~al.}(2000){Ford}, {Kozinsky}, \&
  {Rasio}}]{2000ApJ...535..385F}
{Ford}, E.~B., {Kozinsky}, B., \& {Rasio}, F.~A. 2000, The Astrophysical
  Journal, 535, 385

\bibitem[{{Gevorgyan}(2021)}]{gevorgyan2021}
{Gevorgyan}, Y. 2021, Astronomy \& Astrophysics, 650, A141

\bibitem[{{Guimar{\~a}es} \& {Valio}(2018)}]{2018AJ....156...50G}
{Guimar{\~a}es}, A. \& {Valio}, A. 2018, The Astronomical Journal, 156, 50

\bibitem[{{Hamilton} \& {Burns}(1992)}]{Hamilton}
{Hamilton}, D.~P. \& {Burns}, J.~A. 1992, Icarus, 96, 43

\bibitem[{{Hord} {et~al.}(2021){Hord}, {Col{\'o}n}, {Kostov}, {Galgano},
  {Ricker}, {Vanderspek}, {Seager}, {Winn}, {Jenkins}, {Barclay}, {Caldwell},
  {Essack}, {Fausnaugh}, {Guerrero}, \& {Wohler}}]{2021AJ....162..263H}
{Hord}, B.~J., {Col{\'o}n}, K.~D., {Kostov}, V., {et~al.} 2021, The
  Astronomical Journal, 162, 263

\bibitem[{{Hurford} {et~al.}(2016){Hurford}, {Asphaug}, {Spitale}, {Hemingway},
  {Rhoden}, {Henning}, {Bills}, {Kattenhorn}, \& {Walker}}]{Hurford}
{Hurford}, T.~A., {Asphaug}, E., {Spitale}, J.~N., {et~al.} 2016, Journal of
  Geophysical Research (Planets), 121, 1054

\bibitem[{{Innanen} {et~al.}(1997){Innanen}, {Zheng}, {Mikkola}, \&
  {Valtonen}}]{1997AJ....113.1915I}
{Innanen}, K.~A., {Zheng}, J.~Q., {Mikkola}, S., \& {Valtonen}, M.~J. 1997, The
  Astronomical Journal, 113, 1915

\bibitem[{{Kaula}(1964)}]{Kaula}
{Kaula}, W.~M. 1964, Reviews of Geophysics and Space Physics, 2, 661

\bibitem[{{Leinhardt} {et~al.}(2012){Leinhardt}, {Ogilvie}, {Latter}, \&
  {Kokubo}}]{Leinhardt}
{Leinhardt}, Z.~M., {Ogilvie}, G.~I., {Latter}, H.~N., \& {Kokubo}, E. 2012,
  Monthly Notices of the Royal Astronomical Society, 424, 1419

\bibitem[{{Makarov}(2015)}]{2015ApJ...810...12M}
{Makarov}, V.~V. 2015, The Astrophysical Journal, 810, 12

\bibitem[{{Makarov} {et~al.}(2018){Makarov}, {Berghea}, \&
  {Efroimsky}}]{2018ApJ...857..142M}
{Makarov}, V.~V., {Berghea}, C.~T., \& {Efroimsky}, M. 2018, The Astrophysical
  Journal, 857, 142

\bibitem[{{Murray} \& {Dermott}(1999)}]{MurrayDermott}
{Murray}, C.~D. \& {Dermott}, S.~F. 1999, {Solar System Dynamics} (Cambridge
  University Press)

\bibitem[{{Owens} {et~al.}(2021){Owens}, {de Mooij}, {Watson}, \&
  {Hooton}}]{2021MNRAS.503L..38O}
{Owens}, N., {de Mooij}, E.~J.~W., {Watson}, C.~A., \& {Hooton}, M.~J. 2021,
  Monthly Notices of the Royal Astronomical Society, 503, L38

\bibitem[{{Rambaux} \& {Williams}(2011)}]{2011CeMDA.109...85R}
{Rambaux}, N. \& {Williams}, J.~G. 2011, Celestial Mechanics and Dynamical
  Astronomy, 109, 85

\bibitem[{{Sasaki} {et~al.}(2012){Sasaki}, {Barnes}, \& {O'Brien}}]{Sasaki}
{Sasaki}, T., {Barnes}, J.~W., \& {O'Brien}, D.~P. 2012, The Astrophysical
  Journal, 754, 51

\bibitem[{{Schmitt} \& {Mittag}(2020)}]{2020AN....341..497S}
{Schmitt}, J. H.~M.~M. \& {Mittag}, M. 2020, Astronomische Nachrichten, 341,
  497

\bibitem[{{Tokadjian} \& {Piro}(2020)}]{Tokadjian}
{Tokadjian}, A. \& {Piro}, A.~L. 2020, The Astronomical Journal, 160, 194

\bibitem[{{Tokadjian} \& {Piro}(2022)}]{Tokadjian2}
{Tokadjian}, A. \& {Piro}, A.~L. 2022, The Astrophysical Jopurnal Letters, 929,
  L2

\bibitem[{{Valsecchi, G. B.} {et~al.}(2022){Valsecchi, G. B.}, {Rickman, H.},
  {Morbidelli, A.}, {Wi\'{}sniowski, T.}, {Gabryszewski, R.}, \& {Wajer,
  P.}}]{Valsecchi}
{Valsecchi, G. B.}, {Rickman, H.}, {Morbidelli, A.}, {et~al.} 2022, Astronomy
  and Astrophysics, 667, A91

\bibitem[{{Veras} \& {Ford}(2010)}]{2010ApJ...715..803V}
{Veras}, D. \& {Ford}, E.~B. 2010, The Astrophysical Journal, 715, 803

\bibitem[{Veras {et~al.}(2018)Veras, Georgakarakos, G\"ansicke, \&
  Dobbs-Dixon}]{Veras2018}
Veras, D., Georgakarakos, N., G\"ansicke, B.~T., \& Dobbs-Dixon, I. 2018,
  Monthly Notices of the Royal Astronomical Society, 481, 2180

\bibitem[{Walterova {et~al.}(2022)Walterova, B\v{e}hounkov\'{a}, \&
  Efroimsky}]{Walterova}
Walterova, M., B\v{e}hounkov\'{a}, M., \& Efroimsky, M. 2022, Journal of
  Geophysical Research. Planets, Submitted, https://arxiv.org/abs/2301.02476

\end{thebibliography}

\end{document}